\documentclass[11pt]{article}
\usepackage{amsmath}
\usepackage{amssymb}
\usepackage{amsthm}
\usepackage{graphicx}
\usepackage[margin=1in]{geometry}
\usepackage{setspace}
\usepackage{caption}
\usepackage{booktabs}
\usepackage{multirow}
\usepackage{array}
\usepackage{rotating}

\usepackage{subfig}
\usepackage{hyperref}
\hypersetup{pdftex,colorlinks=true,allcolors=blue}
\usepackage{hypcap}

\usepackage{tikz}
\usetikzlibrary{shapes,arrows}
\usepackage{caption}
\newcommand*{\h}{\hspace{5pt}}

\providecommand{\keywords}[1]{\text{Keywords:} #1}

\title{Approximate Bayesian Bootstrap Procedures to Estimate Multilevel Treatment Effects in Observational Studies with Application to Type 2 Diabetes Treatment Regimens}
\author{Anthony D. Scotina\footnote{Division of Mathematics, Computing, and Statistics, Simmons University, Boston, MA 02115}, \ Andrew R. Zullo\footnote{Department of Health Services, Policy, and Practice, Brown University, Providence, RI 02912}, Robert J. Smith\footnote{Warren Alpert Medical School, Brown University, Providence, RI 02912}, and Roee Gutman\footnote{Department of Biostatistics, Brown University, Providence, RI 02912}}
\date{}

\begin{document}

\maketitle

\begin{abstract}

Randomized clinical trials are considered the gold standard for estimating causal effects. Nevertheless, in studies that are aimed at examining adverse effects of interventions, such trials are often impractical because of ethical and financial considerations. In observational studies, matching on the generalized propensity scores was proposed as a possible solution to estimate the treatment effects of multiple interventions. However, the derivation of point and interval estimates for these matching procedures can become complex with non-continuous or censored outcomes. We propose a novel Approximate Bayesian Bootstrap algorithm that result in statistically valid point and interval estimates of the treatment effects with categorical outcomes. The procedure relies on the estimated generalized propensity scores and multiply imputes the unobserved potential outcomes for each unit. In addition, we describe a corresponding interpretable sensitivity analysis to examine the unconfoundedness assumption. We apply this approach to examines the cardiovascular safety of common, real-world anti-diabetic treatment regimens for Type 2 diabetes mellitus in a large observational database.

\noindent
\end{abstract}

\section{Introduction}

\subsection{Overview}

Cardiovascular disease results in substantial morbidity and is the leading cause of mortality in type 2 diabetes mellitus. Cardiovascular mortality rates among individuals with type 2 diabetes are twice the mortality rates of non-diabetic individuals \cite{preis-09}. There is convincing evidence that blood glucose control in type 2 diabetes reduces the risk of microvascular complications, such as retinopathy, nephropathy, and neuropathy \cite{ukpds-98}. Based on available data, it has been noted that improved glucose control in type 2 diabetes does not necessarily reduce adverse macrovascular cardiovascular outcomes, such as myocardial infarction, stroke, and heart failure \cite{gerstein-08}. Investigation of the potential beneficial or harmful effects of new glucose-lowering medications on the incidence of major cardiovascular events via prospective, randomized controlled trials has been mandated by the FDA as part of their drug approval process starting in 2008 \cite{fda-guidance}. However, comparable trials for older anti-diabetic medications are not available, and insights into cardiovascular outcomes from real-world observational data are limited for both new and older antihyperglycemic medications. 

Metformin has been recommended as a first-line agent in type 2 diabetes because of its low cost, high glucose-lowering efficacy, and low risk of hypoglycemia \cite{inzucchi-15}. There are several classes of antihyperglycemic therapies with distinct mechanisms of action recommended as second-line treatments, generally used in combination with metformin \cite{chamberlain-17}. Sulfonylureas, the oldest oral anti-diabetic agents, result in effective glucose-lowering but are associated with weight gain and an increased risk of hypoglycemia \cite{leonard-17}. Pioglitazone, a member of the thiazolidinedione (TZD) drug class, may offer cardiovascular benefits \cite{dormandy-05, seong-15, strongman-18, chan-18}, although it is recommended that pioglitazone should be avoided in patients with or at risk for heart failure because of its potential to cause fluid retention \cite{ponikowski-16}. For the more recently introduced dipeptidyl peptidase-4 inhibitors (DPP-4i), prospective randomized trials have shown no association between DPP-4i and ischemic events such as myocardial infarction or stroke, but their effects on the risk of heart failure is unclear \cite{scirica-13, white-13, green-15, rosenstock-19}. 

A limited number of bimodal comparison studies have suggested either an increase in cardiovascular events with sulfonylureas or a decrease with metformin or pioglitazone \cite{nissen-08, roumie-12}. However, the effects of each of these and other available non-insulin agents individually or with metformin on cardiovascular events still remains uncertain, and future prospective, randomized trials to address these questions are unlikely to be feasible. 

The literature on causal inference with observational studies has generally been focused on methodology for binary treatments \cite{stuart-10, imbens-15}. Yet, providers and patients frequently must choose among several alternative treatments for type 2 diabetes. The propensity score, which is the probability of receiving treatment conditional on a set of observed covariates, is a common metric used to adjust for observed differences in covariates between two treatment groups in non-randomized observational studies \cite{rosenbaum-83}. For more than two treatment groups, the generalized propensity score (GPS) vector represents the probability of receiving any one of the treatments, conditional on the observed covariates \cite{imbens-00, imai-04}. 

Inverse probability weighting and doubly robust estimators were proposed as possible solutions to estimate causal effects in observational studies \cite{mccaffrey-13, yoshida-17, li-18}. McCaffrey et al. \cite{mccaffrey-13} proposed a method that combines a sandwich estimator with generalized boosted models to obtain point and interval estimates for pairwise average treatment effects. Yoshida et al. \cite{yoshida-17} developed matching weights to estimate pairwise average treatment effects across the entire population. Li and Li \cite{li-18} developed balancing weights to estimate pairwise differences in expectations among the subpopulation with the most overlap in covariates across all treatment groups. Possible limitations of weighting methods are that they are limited to specific estimands (e.g., pairwise differences in expectations) and they may suffer from large biases and large standard errors in finite samples because of extreme weights \cite{lopez-15}. 

Matching algorithms have been proposed as a different approach to estimate causal effects of multiple treatments \cite{lopez-15, rassen-11, yang-16, scotina-19_MA}. Scotina and Gutman \cite{scotina-19_MA} observed that matching on the GPS performed well for five treatments or less, but as the number of interventions increases combining matching with a clustering algorithm resulted in better reduction in initial bias. For multiple treatments, statistically valid matching estimators were developed for continuous outcomes and pairwise differences in expectations \cite{yang-16, scotina-19_ME}. Less attention has been given to other outcome types (e.g., binary, time to event, etc.) and other estimands.

To overcome these limitations, we view causal effects as a missing data problem \cite{rubin-78, gutman-15} and propose a combination of the Approximate Bayesian Bootstrap algorithm with $k$-means clustering on the GPS to estimate the causal effects of multiple treatments. The proposed procedure provides statistically valid estimates for different types of outcomes and can be extended to provide interpretable sensitivity analyses. We implemented the proposed procedure to examine the cardiovascular safety of three antihyperglycemic medications to metformin in patients with type 2 diabetes mellitus.

\subsection{Framework}

For units $i=1,\dots,n$, let $Y_{i}^{obs}$, $\mathbf{X}_{i}$, $W_{i}$, and $T_{iw}$ be the observed outcome, set of $P$ covariates, treatment group identification such that $W_{i}\in\mathcal{W}=\{1,\dots,Z\}$, and an indicator function that is equal to 1 if unit $i$ received treatment $w$ and 0 otherwise. Define $n_{w}=\sum_{i=1}^{n}T_{iw}$ as the number of units receiving treatment $w$, where $\sum_{w=1}^{Z}n_{w}=n$. 

A common assumption when estimating causal effects is the stable unit treatment value assumption (SUTVA) \cite{rubin-80}, which requires that the potential outcomes for one unit are unaffected by the treatment assignment of others, and that there are no multiple versions of treatment within the same treatment group. Assuming SUTVA, the potential outcomes for unit $i$ and treatment $w$ is $Y_{i}(W_{i}=w)=Y_{i}(w)$ and $Y_{i}^{obs}=T_{i1}Y_{i}(1)+\cdots+T_{iZ}Y_{i}(Z)$.

With nominal exposure, commonly used estimands of interest with continuous outcomes are the population average treatment effects between treatment groups $j$ and $k$ for all pairs $\{j,k\}\in\mathcal{W}^{2}$, over the group of units receiving reference treatment $t$ \cite{lopez-15}, 
\begin{equation}
\tau_{jk}^{t}\equiv E\left[Y(j)-Y(k)\mid W=t\right].
\label{tau.jk.t}
\end{equation}
This can be estimated by using the sample average treatment effect on the treated (ATT), 
\begin{equation}
\hat{\tau}_{jk}^{t}\equiv \frac{1}{n_{t}}\sum_{i=1}^{n_{t}}T_{it}\left(Y_{i}(j)-Y_{i}(k)\right).
\label{tau.jk.t.hat}
\end{equation}
However, because only one of the potential outcomes is observed and realized for every unit, an important piece of information needed to estimate $\tau^{t}_{jk}$ and is the assignment mechanism, $P(W=t|X,Y(t))$ \cite{lopez-15, rubin-78}. To estimate $\tau_{jk}^{t}$, two restrictions are commonly made on the assignment mechanism: (i) $P(W=t\mid \mathbf{X}, Y(t))=P(W=t\mid \mathbf{X})\equiv r(t,\mathbf{X})$, and (ii) $0<r(t,\mathbf{X})<1$ for all $t\in\mathcal{W}$ \cite{lopez-15}. These assumptions are also referred to as strong ignorability \cite{imbens-00, imai-04}.

With binary outcomes, $\tau_{jk}^{t}$ and measures the difference $P(Y_i(j)=1\mid W_{i}=t) - P(Y_i(k)=1\mid W_{i}=t)$ among units receiving treatment $t$. Other common estimands include the ratio of these probabilities and their odds ratio \cite{fleiss-03}. 

The generalized propensity score (GPS) vector, $R(\mathbf{X})=(r(1,\mathbf{X}),\dots,r(Z,\mathbf{X}))$, is an extension of the binary propensity score to multiple exposure levels. Under strong ignorability, comparing units with similar $R(\mathbf{X})$ results in well-defined causal effect estimates \cite{imai-04}. Generally, in observational studies $R(\mathbf{X})$ is unknown and only an estimate of it $\hat{R}(\mathbf{X})=\{\hat{r}(1,\mathbf{X}),\ldots,\hat{r}(Z,\mathbf{X})\}$ is available. Estimating $R(\mathbf{X})$ is part of the design phase, and several procedures have been described to estimate it \cite{lopez-15}. Here, we assume that the $R(\mathbf{X})$ have been estimated to the satisfaction of the researcher without observing any outcome data.

To obtain transitive causal effects estimates and to reduce extrapolation to units that have zero probability of receiving one of the treatments, we restrict the analysis only to units ``eligible" to receive all treatments, also referred to as those within the range of common support \cite{lopez-15, dehejia-02}. For nominal exposure, Lopez and Gutman \cite{lopez-15} proposed a rectangular common support region defined by $\hat{r}(w,\mathbf{X})\in(\hat{r}_{min}(w,\mathbf{X}), \hat{r}_{max}(w,\mathbf{X}))$ for all $w\in\mathcal{W}$, where
\begin{align*}
	\hat{r}_{min}(w,\mathbf{X})&=\max\{\min(\hat{r}(w,\mathbf{X}\mid W=1)),\dots,\min(\hat{r}(w,\mathbf{X}\mid W=Z))\}\\
	\hat{r}_{max}(w,\mathbf{X})&=\min\{\max(\hat{r}(w,\mathbf{X}\mid W=1)),\dots,\max(\hat{r}(w,\mathbf{X}\mid W=Z))\}.
\end{align*}

The ATTs for units within this rectangular common support region are defined as
\[
	\tau_{jk}^{t,E}=E[Y(j)-Y(k)\mid W=t, E_{i}=1], 
\]
where
\[
	E_{i}=
	\begin{cases}
	1,&\text{if}\ \hat{r}(w, \mathbf{X}_{i})\in(\hat{r}_{min}(w,\mathbf{X}_{i}), \hat{r}_{max}(w,\mathbf{X}_{i}))\ \text{for all}\ w\in\mathcal{W}\\
	0,&\text{otherwise}
	\end{cases}
\]

\section{Methodology}

\subsection{Matching estimators} \label{matching.estimators}

Matching has been proposed as a method to estimate causal effects with multiple treatments. The goal of matching is to balance the distribution of covariates across all treatment groups. Similarity in covariates between units in different treatment groups is defined using a distance measure \cite{scotina-19_MA}. Examples of such distance measures include the linear GPS:
\[
	|\text{logit}(r(w,\mathbf{X}_{i}))-\text{logit}(r(w,\mathbf{X}_{j}))|,\quad i\neq j\in\{1,\dots,n\},\ w\in\{1,\dots,Z\},
\]
the Euclidean distance between $R(\mathbf{X}_{i})$ and $R(\mathbf{X}_{j})$, 
\[
	\{(R(\mathbf{X}_{i})-R(\mathbf{X}_{j}))^{\text{T}}(R(\mathbf{X}_{i})-R(\mathbf{X}_{j}))\}^{1/2}=\sqrt{\sum_{k=1}^{Z}(r(k,\mathbf{X}_{i})-r(k,\mathbf{X}_{j}))^{2}}, 
\]
and the Mahalanobis distance between two vectors $\mathbf{V}_{i},\mathbf{V}_{j}\in\mathbf{V}$, 
\[
	\left\{(\mathbf{V}_{i}-\mathbf{V}_{j})^{\text{T}}\boldsymbol\Sigma^{-1}(\mathbf{V}_{i}-\mathbf{V}_{j})\right\}^{1/2}, 
\]
where $\boldsymbol\Sigma$ is the covariance matrix of $\mathbf{V}_{i}$ and $\mathbf{V}_{j}$ \cite{lopez-15, yang-16, scotina-19_MA}. 

Using a selected distance measure, a common matching algorithm is 1:$L$ nearest neighbor matching, which selects for unit $i$ in the reference treatment $t$ the $L$ individuals in each of the remaining treatment groups with the smallest distance from unit $i$. When $L=1$, each unit in the reference treatment $t$ is paired with its best match from each of the other treatment groups. Increasing $L$ improves precision by reducing the sampling errors, but this improvement may be hindered by increasing bias that arises from the introduction of poorer matches \cite{scotina-19_MA, ho-07}. 

A different characteristic of matching algorithms is whether units can be used as matches more than once. Matching ``with replacement" can often decrease covariates' bias because non-reference group units that are similar to multiple reference group units can be matched multiple times. However, in some cases matching with replacement may result in only a small number of non-reference units being matched \cite{stuart-10}.

Scotina and Gutman \cite{scotina-19_MA} found that for five or fewer treatments, matching with replacement on the Mahalanobis distance of the logit GPS vector provided the largest reduction in initial covariate bias between treatment groups. This algorithm was defined as LGPSMnc by \cite{scotina-19_MA}. As the number of treatments increases, matching on this distance measure performs worse in terms of reducing initial covariates' bias, because the dimension of the GPS vector increases. With more than five treatments, combining a clustering step with matching on the Mahalanobis distance of the logit GPS resulted in larger reduction in covariates' bias, compared to only matching on the Mahalanobis distance of the logit GPS vector.

Development of inference procedures for causal effect estimation with multiple treatments in matched cohorts has been limited. For matching with replacement, asymptotic point and interval estimators were derived for continuous outcomes and differences in means \cite{yang-16, scotina-19_ME}. However, the performance of these estimators has not been examined for other types of outcomes and estimands. With binary treatment, the bootstrap estimator was found to provide valid inferences when matching without replacement \cite{austin-14}. On the other hand, Abadie and Imbens \cite{abadie-08} showed that for matching with replacement the bootstrap method may be statistically invalid in certain situations. The bootstrap estimator was not examined for multiple treatments.

\subsection{Weighting estimators}

Inverse probability weighting (IPW) is a common approach for estimating causal effects with multiple treatments \cite{mccaffrey-13, yoshida-17, li-18, feng-11}. Li and Li \cite{li-18} proposed a general framework for estimating causal effects using generalized propensity score weighting with multiple treatments. Assuming weak unconfoundedness \cite{imbens-00}, $\tau^{t}_{jk}$ can be estimated as 
\begin{align*}
	\hat{\tau}_{jk}^{t,IPW}&=E[\hat{Y}_{i}(j))-E(\hat{Y}_{i}(k)]\\
	&= \left[\left(\sum_{i=1}^{n}\frac{\mathbb{I}(W_{i}=j)Y_{i}^{obs}r(t,\mathbf{X}_{i})}{r(j,\mathbf{X_{i}})}\right)\times\left(\sum_{i=1}^{n}\frac{\mathbb{I}(W_{i}=j)r(t,\mathbf{X}_{i})}{r(j,\mathbf{X_{i}})}\right)^{-1}\right]\nonumber\\
	& \ \ \ \ -  \left[\left(\sum_{i=1}^{n}\frac{\mathbb{I}(W_{i}=k)Y_{i}^{obs}r(t,\mathbf{X}_{i})}{r(k,\mathbf{X_{i}})}\right)\times\left(\sum_{i=1}^{n}\frac{\mathbb{I}(W_{i}=k)r(t,\mathbf{X}_{i})}{r(k,\mathbf{X_{i}})}\right)^{-1}\right],
\end{align*}
where $r(j,\mathbf{X_{i}})$, $r(k,\mathbf{X_{i}})$, and $r(t,\mathbf{X}_{i})$ are commonly unknown, and replaced with their corresponding estimates $\hat{r}(j,\mathbf{X_{i}})$, $\hat{r}(k,\mathbf{X_{i}})$, and $\hat{r}(t,\mathbf{X}_{i})$. To calculate the sampling variance of this estimate, the sandwich estimator, which takes into account the uncertainty in the estimated propensity score, is commonly implemented \cite{mccaffrey-13, li-18}. 

For binary treatment, Gutman and Rubin \cite{gutman-15, gutman-17} found that IPW is generally statistically valid, but it is susceptible to extreme weights, which could yield erratic causal estimates. This is exacerbated with a large number of treatments and non-normally distributed covariates \cite{lopez-15, scotina-19_ME}.

\subsection{Imputation procedures}

Weighting and matching methods implicitly impute the unobserved potential outcomes. A different approach is to explicitly impute the missing potential outcomes \cite{rubin-78, gutman-15, rubin-07}. The imputation procedures proposed in this paper aim to impute the values of missing potential outcomes in each treatment group. Let $\mathbf{Y}^{obs}_{t}=\{Y_{i}^{obs}: W_{i}=t\}$ be the set of observed potential outcomes for treatment $t$, let $\mathbf{Y}^{mis}_{t,t'}=\{Y_{i}(t'): W_{i}=t, t'\neq t\in\mathcal{W}\}$ be the missing potential outcomes for treatment $t'$, for the set of units which received treatment $t$. A Bayesian approach to missing data imputation samples from the posterior predictive distribution, $P(Y^{mis}_{t,t'}\mid Y^{obs})$.  Let $P(Y_{t}^{obs}\mid\theta)$ be the sampling distribution, where $\theta$ are the parameters governing distribution with prior distribution $P(\theta)$. Assuming that $Y^{mis}_{t,t'}$ is independent from $Y^{obs}$ given $\theta$, then samples for $P(Y^{mis}_{t,t'}\mid Y^{obs})$ can be obtained by first drawing $\theta^{*}$ from its posterior distribution, $P(\theta|Y^{obs}) \propto P(\theta)P(Y^{obs}|\theta)$, and then drawing $\mathbf{Y}^{mis}_{t,t'}$, conditional on $\theta^{*}$ \cite{rubin-07b, gelman-13}. 

\subsubsection{Hot deck imputation}

Hot deck methods were proposed as possible techniques to impute missing data \cite{andridge-10}. These methods replace missing values of one unit (the ``recipient") with observed values from another unit (the ``donor"). Two common implementations of hot deck imputation methods are the random hot deck and the deterministic hot deck imputation. In the random hot deck imputation, the donor is randomly selected from a set of potential donors, where ``potential" is defined according to a distance metric. In the deterministic hot deck imputation, a single donor is identified using a nearest neighbor approach based on a distance metric \cite{andridge-10, tu-13}. 

The distance measures that were described in Section~\ref{matching.estimators} can be used to create donor pools for hot deck methods. For imputing missing potential outcomes when comparing multiple treatments, one possible procedure to create donor pools is based on $k$-means clustering \cite{tu-13}. This procedures creates donor pools using  $k$-means clustering on the logit transformation of $\hat{R}(\mathbf{X})$, such that patients within each cluster are roughly similar on each component of $\text{logit}(\hat{R}(\mathbf{X}))=(\text{logit}(\hat{r}(1,\mathbf{X})),\dots,\text{logit}(\hat{r}(Z,\mathbf{X})))$, and there is at least one patient from each treatment group in each cluster. Imputation is then performed within each cluster separately. A limitation of this clustering approach is that with a large number of clusters, one or more treatment groups may have a small number or no individuals within a cluster, which would require extrapolation to that cluster \cite{lopez-15}. Another limitation of hot deck imputation is that it is not a proper Bayesian imputation method, which results in an underestimation of the sampling variance \cite{andridge-10, rubin-86}.

\subsubsection{Approximate Bayesian Bootstrap multiple imputation}

The Bayesian Bootstrap was proposed as an imputation method, which is based on observed values and reflects the uncertainty in the estimation of population parameters \cite{rubin-86}. For a categorical outcome variable, let $\theta=(\theta_{1},\dots,\theta_{L})$ represent the probabilities of the outcome taking values $d_{1},\dots,d_{L}$. Assuming the prior distribution $P(\theta) \propto \prod_{l=1}^{L} \theta_{l}^{-1}$, then the posterior distribution is $P(\theta|Y^{obs}_{t}) \propto \prod_{l=1}^{L} \theta^{q_l-1}$, where $q_{l}$ is the number of times that the value $d_{l}$ appears in $Y^{obs}_{t}$. The Bayesian Bootstrap algorithm draws $\theta^*$ from $P(\theta|Y^{obs}_{t})$, then $Y^{mis}_{t,t'}$ are drawn independently from $d_{1},\dots,d_{l}$ with probability $\theta^*_{1},\ldots,\theta^*_{l}$ \cite{rubin-86}. The Approximate Bayesian Bootstrap (ABB) was proposed as computationally simpler approximation of the Bayesian Bootstrap \cite{rubin-86}. For treatments $t$ and $t'\neq t$, ABB imputes the missing potential outcomes $\mathbf{Y}^{mis}_{t,t'}$ for treatment group $t$ by: (i) drawing $n_{t'}$ components with replacement from $\mathbf{Y}^{obs}_{t'}$; (ii) drawing $n_{t}$ components with replacement from the $n_{t'}$ deaws in (i). This double resampling procedure approximates the Bayesian Bootstrap and results in valid statistical inference \cite{rubin-81}. Both the Bayesian Bootstrap and the ABB do not adjust for potential covariates. Lavori \cite{lavori-95} proposed a procedure that performs the ABB within propensity score subclasses.  

We propose a procedure that combines $k$-means clustering with multiple imputation using ABB to estimate causal effects of multiple treatments. Formally, to estimate $\tau_{jk}^{t}$:
\begin{enumerate}
	\item Partition patients into $Q$ subclasses based on $\text{logit}(\hat{R}(\mathbf{X}))$ using $k$-means clustering, and ensure that there is at least one treated unit from each treatment group in each cluster. 
	\item Within each cluster $q\in\{1,\dots,Q\}$:
	\begin{enumerate}
		\item Let $O_{w}^{q}$ be the set of units in subclass $q$ which received treatment $w$, and let $n_{w}^{q}=|O_{w}^{q}|$ be the cardinality of $O_{w}^{q}$. For each treatment $w\neq t\in\mathcal{W}$, draw $n_{w}^{q}$ values from $O_{w}^{q}$ with replacement, where each element has $1/n_{w}^{q}$ probability of being selected. This forms the donor pool, $\tilde{O}_{w}^{q}$, for each treatment $w$. 
		\item For each $w$, draw $n_{t}^{q}$ values from $\tilde{O}_{w}^{q}$ with replacement. Each of the elements in $\tilde{O}^{q}_{w}$ has equal probability of being selected. 
	\end{enumerate}
	\item Repeat steps 2(a) and 2(b) $M=25$ times. 
\end{enumerate}
This creates $M$ imputed datasets, each containing imputed missing potential outcomes for patients in treatment group $t$. 

Let $\hat{\tau}_{jk}^{t(m)}$ denote the estimated ATT between treatments $j$ and $k$ from imputed dataset $m=1,\dots,M$, and let $\hat{V}_{jk}^{(m)}$ be its estimated sampling variance. The point estimate for $\tau_{jk}^{t}$ is
\[
	\hat{\tau}_{jk}^{t}=\frac{1}{M}\sum_{m=1}^{M}\hat{\tau}_{jk}^{t(m)}, 
\]
and the standard error is estimated by
\[
	\sqrt{\frac{1}{M-1}\sum_{m=1}^{M}\left(\hat{\tau}_{jk}^{t(m)}-\hat{\tau}_{jk}\right)^{2}+\frac{1}{M}\sum_{m=1}^{M}\hat{V}_{jk}^{(m)}}.
\]
While we use $M=25$ imputations, the standard error can be estimated using percentiles of $\hat{\tau}_{jk}^{t(m)}$ when $M>100$ \cite{gutman-15}. 

One limitation of implicit imputation methods such as weighting and matching is that inference is generally limited to specific estimands. Obtaining interval estimates for other estimands is generally complex and relies on asymptotic approximations. An advantage of the proposed ABB procedure is that it can provide valid inference for any estimand, because it explicitly imputes the missing potential outcomes.  

\section{Simulation Study}

\subsection{Design}

To examine the operating characteristics of the different procedures, we constructed simulation analyses. The first set of factors describes the covariate distributions of $Z=3$ treatment groups, which are either known to the investigator, or can be estimated without outcome data. The second set of factors describes the response surfaces, which are unknown to the investigator.  

We generated the $P=18$ covariates' values for $n_{1}=1200$, $n_{2}=2400$, and $n_{3}=4800$ units receiving treatments 1, 2, and 3, respectively, from multivariate skew-$t$ distributions with 7 degrees of freedom \cite{azzalini-03}:
  \begin{align*}
\mathbf{X}_{i}\mid\{W_{i}=1\}&\sim \text{Skew-}t_{7,18}(\boldsymbol\mu_{1}, \mathbf{I}_{18}, \boldsymbol\eta),\ i=1,\dots,n_{1},\\
\mathbf{X}_{i}\mid\{W_{i}=2\}&\sim \text{Skew-}t_{7,18}(\boldsymbol\mu_{2}, \mathbf{I}_{18}, \boldsymbol\eta),\ i=n_{1}+1,\dots,n_{1}+ n_{2},\\
\mathbf{X}_{i}\mid\{W_{i}=3\}&\sim \text{Skew-}t_{7,18}(\boldsymbol\mu_{3}, \mathbf{I}_{18}, \boldsymbol\eta),\ i=n_{1}+n_{2}+1,\dots, n_{1}+n_{2}+n_{3},
\end{align*}
where  $\boldsymbol\mu_{w}=\text{vec}(1_{P}\otimes b_{w})$, where $1_{P}$ is a $P\times1$ vector of 1s, and $b_{w}$ is a $3\times1$ vector such that the $w$ value is equal to $b$ and the rest are zeros, $b\in\{0, 0.25, 0.50, 0.75, 1\}$. As the value of $b$ increases there is larger initial bias across the different treatment groups.

We assumed that $Y_i(w) \in \{d_1,\ldots,d_{L}\}$, and we only examined monotone response surfaces. Binary potential outcomes, $Y_{i}(w) \in \{0,1\}$, were simulated from Bernoulli distributions with success probability $G^{-1}_{1}(\mathbf{X}_{i}\boldsymbol\beta_{X}')$ in reference treatment group 1, $G^{-1}_{2}(\gamma\mathbf{X}_{i}\boldsymbol\beta_{X}')$ in treatment group 2, and $G^{-1}_{3}(\gamma\mathbf{X}_{i}\boldsymbol\beta_{X}')$ in treatment group 3, where $\boldsymbol\beta_{X}=(2,4,6,1,\dots,1)$ is a $P\times1$ vector that governs the association between $\mathbf{X}_{i}$ and $Y_{i}(w)$. When $\gamma=-1$ the response surfaces for treatment groups 2 and 3 are mirror images of the response surface in treatment group 1 multiplied by a constant. We used two link functions for monotone response surfaces: the logistic, and the normal (probit), both of which are symmetric. 

For ordinal $Y_{i}(w)\in\{1,2,3,4,5\}$, the values were simulated using the proportional odds model,
\[
	\text{logit}[P(Y_{i}(w)\leq j \mid \mathbf{X}_{i})]=\alpha_{j}+\mathbf{X}_{i}\boldsymbol\beta_{X}'\qquad j=1,\dots,4,
                                                                                                                                                                                                                                                                                                                                                                                                             \]
where $\alpha_{1}=1$, $\alpha_{2}=0.05$, $\alpha_{3}=-0.05$, and $\alpha_{4}=-1$. 

The simulations were performed as a full factorial design of the factors described in Table~\ref{sim.factors}. Each simulation configuration was replicated 200 times. At each replication $\ell$, $\ell=1,\dots,200$, we estimated each pairwise ATT and the standard error, for each of the procedures. Using these values we recorded at each replication the following measures:
 \begin{align*}
bias_{\ell(j,k)}&=\hat{\tau}_{\ell(j,k)}^{t}-\tau_{\ell(j,k)}^{t}\\
coverage_{\ell(j,k)}&=\left\{1\ \text{if}\ \tau_{\ell(j,k)}^{t}\in\hat{\tau}_{\ell(j,k)}^{t}\pm1.96\times\text{SE}(\hat{\tau}_{\ell(j,k)}^{t}),\ 0\ \text{otherwise}\right\}
\end{align*}

\begin{table}[!htb]
\centering
\caption{Simulation factors}
\begin{tabular}{l l}
\toprule
Factor & Levels of factor\\
\midrule
$b$ & $\{0, 0.25, 0.50, 0.75, 1.00\}$\\
$\eta$ & $\{-3.5,0,3.5\}$\\
$\gamma$ & $\{-1,1\}$\\
$G_w$ & $\{\text{logistic},\text{probit}\}$\\
\bottomrule
\end{tabular}
\label{sim.factors}
\end{table}

To summarize the bias, we calculated the mean and standard deviation of $|bias_{\ell(j,k)}|$ across the 200 replications. The median $SE(\hat{\tau}^t_{\ell(j,k)})$ across the 200 replications was used to summarize the standard errors of each pairwise comparison. 

We summarized matching performance using maximum absolute standardized pairwise bias, $Max2SB_{p}$, for each pair of treatments and each covariate $p$, 
\[
  Max2SB_{p}=\max\left(|SB_{p12}|, |SB_{p13}|, |SB_{p23}|, \dots\right),
  \]
This metric reflects the largest discrepancy in estimated means between any two treatment groups for covariate $p$ \cite{mccaffrey-13, lopez-15, scotina-19_MA}. McCaffrey et al. \cite{mccaffrey-13} advocated a cutoff of 0.20 but maintained that larger cutoffs may be appropriate for different applications. We summarized this metric over all covariates for each replication by calculating the maximum bias, $MaxMax2SB=\max_{p=1,\dots,P}(Max2SB_{p})$. To summarize clustering performance, the $SB_{pjk}$ were calculated within each cluster and weighted across clusters, with weights proportional to the number of units within each cluster. 

The following methods were compared: nearest neighbor matching for multiple treatments using the LGPSMnc algorithm from Scotina and Gutman \cite{scotina-19_MA}, IPW \cite{mccaffrey-13}, and ABB multiple imputation with $k$-means clustering using $Q\in\{1,3,5,7\}$ clusters. 

\subsection{Results for clustering and matching performance}

\begin{figure}[!ht]
\centering
\includegraphics[scale=0.5]{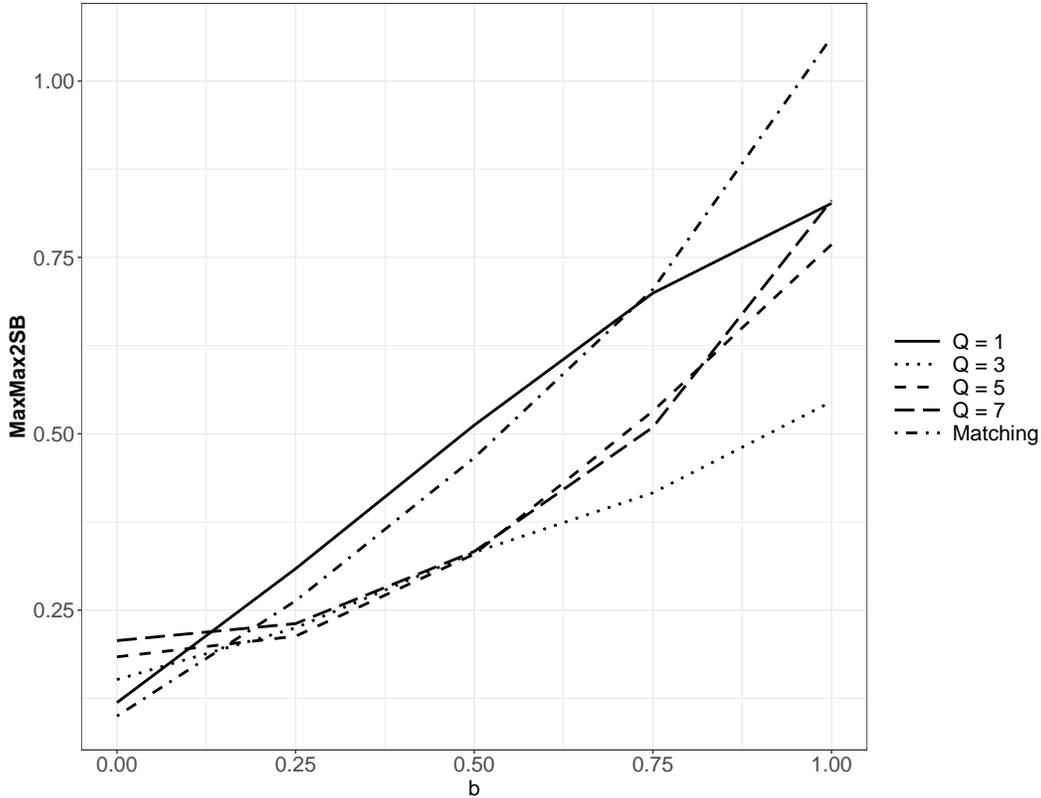}
\caption{$MaxMax2SB$ across different levels of $b$ for ABB and matching}
\label{maxmax2sb.line}
\end{figure}

Covariates' balance across treatment groups for $k$-means clustering and matching is depicted in Figure~\ref{maxmax2sb.line}. For each level of $Q$ and matching, $MaxMax2SB$ increases with $b$ and it is largest for $b=1.00$. This trend is the strongest for matching, where its largest $MaxMax2SB$ is 1.06 for $b=1.00$. ABB with $k$-means clustering with $Q=3$ was most resistant to an increase in $b$, where its $MaxMax2SB$ of 0.55 for $b=1$ was the smallest of the examined procedures. For small values of $b$ matching had the smallest $MaxMax2SB$.

\subsection{Results for 95\% interval coverage, bias, and standard error}

To identify the simulations' factor with the largest influence on the performance of the different estimation methods, we examined the factors' MSE for interval coverage between treatments 1 and 2 and treatments 1 and 3 (as in the works of Lopez and Gutman, Scotina and Gutman, and Cangul et al. \cite{lopez-15, scotina-19_MA, cangul-09}). Across all of the methods that were examined, initial covariates' bias, $b$, was found to be the most influential factor on coverage rates. 

We will begin by describing the simulation results for $Y_{i}(w) \in \{0,1\}$. Tables~\ref{sim.null} and \ref{sim.nonnull} show the 95\% coverage rates, mean absolute bias, and standard errors for the ATT measured as pairwise risk differences with treatment group 1 as the reference across levels of $b$ when the treatment effect is null or non-null, respectively. When there is small initial bias, all of the methods perform well with IPW having the smallest standard errors. However, as the initial bias increases, ABB with either 5 or 7 clusters generally has the smallest absolute bias and standard errors. ABB with $Q=7$ clusters generally results smaller average absolute bias for $b \leq 0.5$ compared to $Q=5$, while ABB with $Q=5$ results in smaller average absolute biases for $b >0.5$. In addition, ABB with $Q=5$ yields coverages that are at or above nominal for both null and non-null treatment effects for each level of $b$. ABB with $Q=1$ cluster is only valid when $b<0.25$. For $b\geq0.50$, IPW yielded the largest mean absolute bias and standard error, which results in coverage rates that are lower than nominal. For $b\leq0.50$, matching estimators yield nominal coverage rates, with mean absolute biases and standard errors that are similar to those obtained using ABB with $Q\in\{3,5,7\}$. For $b\geq0.75$, matching estimator coverage rates are below nominal, and the mean absolute biases and standard errors are larger than the ones observed for ABB with $Q \in \{5,7\}$ clusters.

\begin{sidewaystable}[!ht]
\small
\centering
\caption{Simulated coverage, bias, and standard deviation of bias of ATT estimators across levels of $b$, when treatment effect is null}
\begin{tabular}{l l c c c c c c c}
\toprule
& & \multicolumn{7}{c}{Treatment Group Comparison} \\
\cline{3-9}\\
& & \multicolumn{3}{c}{1 vs. 2} & & \multicolumn{3}{c}{1 vs. 3}\\
\cline{3-5}  \cline{7-9} \\
$b$ & Method & Coverage & Mean absolute bias (SD) & Std. Error && Coverage & Mean absolute bias (SD) & Std. Error \\
0.00 & \underline{ABB} & & & & & & & \\
& $Q=1$ & 0.99 & 0.0146 (0.0139) & 0.0270 && 1.00 & 0.0123 (0.0097) & 0.0261 \\
& $Q=3$ & 1.00 & 0.0109 (0.0081) & 0.0269 && 1.00 & 0.0095 (0.0074) & 0.0259 \\
& $Q=5$ & 1.00 & 0.0118 (0.0092) & 0.0273 && 1.00 & 0.0104 (0.0073) & 0.0259 \\
& $Q=7$ & 1.00 & 0.0098 (0.0078) & 0.0266 && 1.00 & 0.0083 (0.0072) & 0.0259 \\ 
& \underline{Matching} & 0.99 & 0.0155 (0.0116) & 0.0228 && 0.97 & 0.0165 (0.0120) & 0.0215 \\
& \underline{IPW} & 1.00 & 0.0093 (0.0068) & 0.0178 && 1.00 & 0.0107 (0.0070) & 0.0162 \\
\midrule
0.25 & \underline{ABB} & & & & & & & \\
& $Q=1$ & 0.97 & 0.0230 (0.0146) & 0.0265 && 0.74 & 0.0410 (0.0163) & 0.0254 \\
& $Q=3$ & 1.00 & 0.0179 (0.0126) & 0.0275 && 0.96 & 0.0238 (0.0121) & 0.0257 \\
& $Q=5$ & 1.00 & 0.0109 (0.0079) & 0.0279 && 1.00 & 0.0191 (0.0101) & 0.0259 \\ 
& $Q=7$ & 1.00 & 0.0155 (0.0111) & 0.0278 && 1.00 & 0.0109 (0.0090) & 0.0263 \\
& \underline{Matching} & 0.97 & 0.0154 (0.0127) & 0.0250 && 0.99 & 0.0156 (0.0106) & 0.0227 \\
& \underline{IPW} & 0.99 & 0.0129 (0.0083) & 0.0210 && 0.99 & 0.0100 (0.0073) & 0.0179 \\
\midrule
0.50 & \underline{ABB} & & & & & & & \\
& $Q=1$ & 0.60 & 0.0441 (0.0166) & 0.0254 && 0.01 & 0.0772 (0.0145) & 0.0239 \\
& $Q=3$ & 0.99 & 0.0258 (0.0150) & 0.0285 && 0.68 & 0.0427 (0.0150) & 0.0255 \\
& $Q=5$ & 1.00 & 0.0155 (0.0121) & 0.0297 && 0.99 & 0.0213 (0.0125) & 0.0281 \\
& $Q=7$ & 0.99 & 0.0169 (0.0128) & 0.0307 && 1.00 & 0.0158 (0.0117) & 0.0279 \\
& \underline{Matching} & 0.96 & 0.0250 (0.0194) & 0.0325 && 0.88 & 0.0247 (0.0204) & 0.0270 \\
& \underline{IPW} & 0.94 & 0.0656 (0.0669) & 0.0559 && 0.86 & 0.0697 (0.0603) & 0.0533 \\
\midrule
0.75 & \underline{ABB} & & & & & & & \\
& $Q=1$ & 0.45 & 0.0466 (0.0160) & 0.0233 && 0.00 & 0.0866 (0.0136) & 0.0216 \\
& $Q=3$ & 0.95 & 0.0246 (0.0168) & 0.0294 && 0.71 & 0.0373 (0.0194) & 0.0255 \\
& $Q=5$ & 0.99 & 0.0257 (0.0190) & 0.0376 && 0.99 & 0.0166 (0.0116) & 0.0311 \\
& $Q=7$ & 0.94 & 0.0328 (0.0216) & 0.0381 && 1.00 & 0.0226 (0.0154) & 0.0338 \\
& \underline{Matching} & 0.82 & 0.0556 (0.0448) & 0.0506 && 0.84 & 0.0442 (0.0306) & 0.0386 \\
& \underline{IPW} & 0.70 & 0.1625 (0.1388) & 0.1057 && 0.75 & 0.1672 (0.1638) & 0.0894 \\
\midrule
1.00 & \underline{ABB} & & & & & & & \\
& $Q=1$ & 0.14 & 0.0590 (0.0149) & 0.0215 && 0.00 & 0.1074 (0.0138) & 0.0191 \\
& $Q=3$ & 0.95 & 0.0281 (0.0226) & 0.0382 && 0.97 & 0.0210 (0.0152) & 0.0295 \\
& $Q=5$ & 0.94 & 0.0458 (0.0346) & 0.0524 && 0.97 & 0.0308 (0.0260) & 0.0391 \\
& $Q=7$ & 0.93 & 0.0489 (0.0364) & 0.0519 && 0.88 & 0.0551 (0.0348) & 0.0516 \\
& \underline{Matching} & 0.79 & 0.0954 (0.0695) & 0.0816 && 0.93 & 0.0534 (0.0439) & 0.0603 \\
& \underline{IPW} & 0.54 & 0.2342 (0.2086) & 0.1087 && 0.51 & 0.3210 (0.2786) & 0.0970 \\
\bottomrule
\end{tabular}
\label{sim.null}
\end{sidewaystable}


\begin{sidewaystable}[!ht]
\small
\centering
\caption{Simulated coverage, bias, and standard deviation of bias of ATT estimators across levels of $b$, when treatment effect is non-null}
\begin{tabular}{l l c c c c c c c}
\toprule
& & \multicolumn{7}{c}{Treatment Group Comparison} \\
\cline{3-9}\\
& & \multicolumn{3}{c}{1 vs. 2} & & \multicolumn{3}{c}{1 vs. 3}\\
\cline{3-5}  \cline{7-9} \\
$b$ & Method & Coverage & Mean absolute bias (SD) & Std. Error && Coverage & Mean absolute bias (SD) & Std. Error \\
0.00 & \underline{ABB} & & & & & & & \\
& $Q=1$ & 1.00 & 0.0119 (0.0108) & 0.0271 && 1.00 & 0.0129 (0.0091) & 0.0260 \\
& $Q=3$ & 1.00 & 0.0114 (0.0081) & 0.0266 && 1.00 & 0.0092 (0.0076) & 0.0261 \\
& $Q=5$ & 1.00 & 0.0104 (0.0079) & 0.0264 && 1.00 & 0.0094 (0.0071) & 0.0257 \\
& $Q=7$ & 1.00 & 0.0098 (0.0076) & 0.0269 && 1.00 & 0.0077 (0.0068) & 0.0259 \\ 
& \underline{Matching} & 0.99 & 0.0143 (0.0105) & 0.0227 && 0.99 & 0.0127 (0.0090) & 0.0214 \\
& \underline{IPW} & 1.00 & 0.0095 (0.0070) & 0.0177 && 1.00 & 0.0081 (0.0066) & 0.0162 \\
\midrule
0.25 & \underline{ABB} & & & & & & & \\
& $Q=1$ & 1.00 & 0.0154 (0.0118) & 0.0270 && 0.80 & 0.0362 (0.0157) & 0.0256 \\
& $Q=3$ & 1.00 & 0.0167 (0.0105) & 0.0281 && 1.00 & 0.0195 (0.0111) & 0.0260 \\
& $Q=5$ & 1.00 & 0.0141 (0.0101) & 0.0276 && 0.99 & 0.0174 (0.0121) & 0.0262 \\ 
& $Q=7$ & 1.00 & 0.0112 (0.0084) & 0.0282 && 1.00 & 0.0132 (0.0085) & 0.0262 \\
& \underline{Matching} & 0.96 & 0.0164 (0.0144) & 0.0252 && 0.99 & 0.0150 (0.0105) & 0.0228 \\
& \underline{IPW} & 1.00 & 0.0131 (0.0098) & 0.0209 && 1.00 & 0.0091 (0.0070) & 0.0179 \\
\midrule
0.50 & \underline{ABB} & & & & & & & \\
& $Q=1$ & 0.60 & 0.0439 (0.0180) & 0.0254 && 0.02 & 0.0775 (0.0153) & 0.0243 \\
& $Q=3$ & 0.89 & 0.0339 (0.0173) & 0.0284 && 0.72 & 0.0426 (0.0133) & 0.0262 \\
& $Q=5$ & 0.98 & 0.0268 (0.0136) & 0.0298 && 0.98 & 0.0244 (0.0128) & 0.0281 \\
& $Q=7$ & 1.00 & 0.0151 (0.0124) & 0.0315 && 0.99 & 0.0162 (0.0112) & 0.0282 \\
& \underline{Matching} & 0.94 & 0.0246 (0.0192) & 0.0327 && 0.98 & 0.0178 (0.0133) & 0.0269 \\
& \underline{IPW} & 0.83 & 0.0956 (0.0811) & 0.0802 && 0.84 & 0.0813 (0.0780) & 0.0555 \\
\midrule
0.75 & \underline{ABB} & & & & & & & \\
& $Q=1$ & 0.39 & 0.0528 (0.0171) & 0.0243 && 0.00 & 0.1022 (0.0144) & 0.0221 \\
& $Q=3$ & 0.95 & 0.0266 (0.0173) & 0.0304 && 0.59 & 0.0454 (0.0172) & 0.0260 \\
& $Q=5$ & 0.99 & 0.0248 (0.0173) & 0.0378 && 1.00 & 0.0192 (0.0140) & 0.0325 \\
& $Q=7$ & 0.96 & 0.0283 (0.0235) & 0.0398 && 0.99 & 0.0235 (0.0182) & 0.0348 \\
& \underline{Matching} & 0.97 & 0.0397 (0.0306) & 0.0495 && 0.89 & 0.0434 (0.0280) & 0.0388 \\
& \underline{IPW} & 0.49 & 0.2627 (0.1706) & 0.1036 && 0.57 & 0.2138 (0.1734) & 0.1147 \\
\midrule
1.00 & \underline{ABB} & & & & & & & \\
& $Q=1$ & 0.14 & 0.0627 (0.0143) & 0.0223 && 0.00 & 0.1038 (0.0123) & 0.0197 \\
& $Q=3$ & 0.97 & 0.0307 (0.0207) & 0.0380 && 1.00 & 0.0216 (0.0145) & 0.0294 \\
& $Q=5$ & 0.96 & 0.0496 (0.0314) & 0.0528 && 0.97 & 0.0328 (0.0251) & 0.0421 \\
& $Q=7$ & 0.91 & 0.0595 (0.0454) & 0.0577 && 0.96 & 0.0348 (0.0235) & 0.0478 \\
& \underline{Matching} & 0.88 & 0.0762 (0.0658) & 0.0788 && 0.73 & 0.0974 (0.0625) & 0.0630 \\
& \underline{IPW} & 0.48 & 0.2294 (0.1770) & 0.1064 && 0.52 & 0.3045 (0.2337) & 0.1212 \\
\bottomrule
\end{tabular}
\label{sim.nonnull}
\end{sidewaystable}

Because ABB with $k$-means clustering is not restricted to a specific estimand, we examined its performance for estimating the log odds ratio and the log risk ratios. Tables~\ref{sim.OR.null}--\ref{sim.RR.nonnull} display the 95\% coverage, mean absolute bias and standard errors across the different levels of $b$. The results are generally similar to the ones observed for risk differences. Specifically, ABB with $Q\in\{5,7\}$ clusters generally has coverages that are close to nominal. Both methods may result in undercoverage when the initial bias is large.

The results for ordinal outcomes are similar to the ones observed for binary outcomes. Table~\ref{sim.ordinal} shows the 95\% coverage rates, mean absolute bias, and standard errors for the ATT measured as pairwise risk differences with treatment group 1 as the reference across levels of $b$ when the outcome is ordinal. Each of the examined methods perform well with small initial bias. ABB with either 5 or 7 clusters generally yield the smallest absolute bias and standard errors as the initial bias increases. 

\section{Cardiovascular safety of second-line non-insulin antihyperglycemic therapy}

\subsection{Data description}

The Clinical Practice Research Datalink (CPRD), established in 1987, is a longitudinal, anonymized database which includes more than 13 million people enrolled from over 600 general practitioners in the United Kingdom (U.K.). Available data include basic patient demographics and registration details, medical history events including symptoms, signs and diagnoses, clinical test data, and details of all issued prescriptions. In addition, the general practitioners' practice postcodes and eligible patient residence postcodes are linked to neighborhood characteristics to obtain measures of area level deprivation and a rural-urban classification.

The CPRD represents complete data from the U.K. National Healthcare System, and previous validation studies have reported on the accuracy of its diagnostic data \cite{khan-10}. Diagnostic information in the CPRD is coded using Read codes, the standard clinical terminology system used in the U.K. A subset of patients is linked to additional datasets from the Hospital Episode Statistics inpatient data, which include information about National Health Service inpatient visits. To ensure that we have complete hospitalizations records we restricted the analysis only to the subset of patients that can be linked. Primary and additional causes of hospital admissions are coded using ICD-10 codes. 

We restrict our analysis to CPRD patients who were new users of antihyperglycemic treatments. Specifically, to be included in the study patients had (i) received their first ever prescription for a second-line therapy for type 2 diabetes mellitus between January 2007 and December 2013, and used it for 60 days; (ii) had at least six months of continuous follow-up prior to the cohort entry date; (iii) had a diagnosis of type 2 diabetes mellitus at anytime prior to the date of initiation of the second-line therapy. The index date was defined as 60 days after the first filled prescription for the second-line agent, and patients were grouped based on their second-line antihyperglycemic regimen: sulfonylurea (SU), dipeptidyl peptidase-4 (DPP-4) inhibitor, or thiazolidinedione (TZD). Because the majority of patients initiated dual-therapy with gliclazide (G) (90\% of SU users), sitagliptin (S) (72\% of DPP-4i users), or pioglitazone (P) (83\% of TZD users), we further restricted the study population to patients who initiated dual-therapy with one of these three agents. The analyses were performed using an intention-to-treat approach, such that patients were assigned to a cohort once they met the criteria, regardless of future changes in treatment regimen not otherwise specified, such as adding a third agent (see Appendix for a summary of cohort selection).

Covariates included demographics (e.g., age, sex, race, etc.), clinical measurements (e.g., HbA1c, BMI, LDL cholesterol, etc.), comorbidities, concurrent medication use, year of the index date, diabetes duration until the index date, and time on metformin before add-on initiation. Duration of diabetes was calculated from the date of first observed diabetes claim to the index date, and time on metformin was calculated from the date of first metformin use to the index date. Comorbidities were defined if they were diagnosed within two years prior to the index date, and concurrent medication use was defined as at least one filled prescription in the year prior to the index date. 

The primary analysis was the average treatment effect among gliclazide add-on users, estimated using three-year pairwise risk differences of major adverse cardiovascular events (MACE) and all-cause mortality (ACM). MACE included fatal/non-fatal CVD including myocardial infarction, fatal/non-fatal CHF, and stroke. These outcome events were ascertained using ICD-10 codes for patients linked with Hospital Episode Statistics (see Appendix for ICD-10 codes) and Read codes for the rest \cite{ekstrom-16}. 

Table~\ref{baseline.characteristics.support} lists the covariates, along with a $p$-value testing the null hypothesis of no dependency between second-line treatment initiation and each covariate. To estimate the GPS model, we used a multinomial logistic regression model that included all of the baseline covariates in Table~\ref{baseline.characteristics.support} as explanatory variables. Causal effects were estimated using ABB with $k$-means clustering on $\text{logit}(\hat{R}(\mathbf{X}))$ with $Q\in\{1,3,5,7\}$ clusters, 1:1:1 nearest neighbor matching on the Mahalanobis distance of the logit GPS vector with replacement \cite{scotina-19_MA}, and IPW \cite{mccaffrey-13}. Because some of the baseline covariates had missing values, we created 20 multiple complete datasets by applying the fully conditional specification approach in each arm separately using the mice package \cite{su-11}. We restricted the analysis only to units within the range of common support, $E_{i}$. IPW and matching were implemented in each imputed dataset and the results were combined using the common multiple imputation rules \cite{rubin-87}. When using ABB, imputation was performed in each of the complete datasets and final point and interval estimates were obtained using the common combination rules for two-stage imputation \cite{harel-07}. 

Of the 15,484 metformin users included in the analysis, 10,487 (68\%) had initiated dual-therapy with gliclazide, 3,180 (21\%) with sitagliptin and 1,817 (12\%) with pioglitazone. The median patient age was between 60 and 61 years for each group. Diabetes duration was less than five years for 82\% of the patients and between five and ten years for 18\% of the patients. Median time on metformin monotherapy was approximately 2.5-2.6 years for all groups. HbA1c levels were between 8.4-8.7 for each group, reflecting similar blood glucose control. Pioglitazone users had the highest prevalence of hypertension, neuropathy, and retinopathy at initiation. 

\begin{table}
\footnotesize
\centering
\caption{Baseline characteristics for patients within the region of common support, cohorts defined by second-line treatment allocation}
\begin{tabular}{l l l l l l l l l l}
\toprule
Variable &&& Gliclazide && Sitagliptin && Pioglitazone && $p$-value\\
\midrule
\textbf{No. of patients} &&& 10487 && 3180 && 1817 &&\\
\textbf{Age, years} &&& 61 (11) && 60 (11) && 60 (12) && $<0.001$\\
\textbf{Women}, $n$ (\%) &&& 3990 (38.0) && 1202 (37.8) && 620 (34.1) && $0.006$\\
\textbf{Race}, $n$ (\%) &&& && && && $0.266$\\
\ \ \ \ White &&& 9522 (90.8) && 2891 (90.9) && 1669 (91.9) &&\\
\ \ \ \ Black &&& 181 (1.7) && 44 (1.4) && 21 (1.2) &&\\
\ \ \ \ Hispanic/other &&& 784 (7.5) && 245 (7.7) && 127 (7.0) &&\\
\textbf{Deprivation index} &&& 3 (2) && 3 (2) && 3 (2) && $0.007$\\
\textbf{Smoking}, $n$ (\%) &&& && && && \\
\ \ \ \ Smoker &&& 2335 (18.2) && 717 (16.7) && 353 (18.3) && $0.144$\\
\textbf{Year, index date}, $n$ (\%) &&& && &&  &&\\
\ \ \ \ 2007 &&& 979 (9.3) && 28 (0.9) && 276 (15.2) &&\\
\ \ \ \ 2008 &&& 1441 (13.7) && 141(4.4) && 403 (22.2) &&\\
\ \ \ \ 2009 &&& 1933 (18.4) && 380 (11.9) && 407 (22.4) &&\\
\ \ \ \ 2010 &&& 1840 (17.5) && 738 (23.2) && 360 (19.8) &&\\
\ \ \ \ 2011 &&& 1642 (15.7) && 717 (22.5) && 244 (13.4) &&\\
\ \ \ \ 2012 &&& 1504 (14.3) && 691 (21.7) && 75 (4.1) &&\\
\ \ \ \ 2013 &&& 1148 (10.9) && 485 (15.3) && 52 (2.9) &&\\
\textbf{Diabetes duration}, years &&& 3.3 (1.8) && 3.7 (1.9) && 3.2 (1.5) && $<0.001$\\
\textbf{Time on metformin}, years &&& 2.5 (1.9) && 2.6 (2.0) && 2.6 (2.0) && $0.027$\\
\textbf{HbA1c}, \% &&& 8.7 (1.6) && 8.4 (1.5) && 8.5 (1.5) && $<0.001$\\
\textbf{BMI}, kg/m$^{2}$ &&& 32.1 (5.4) && 32.7 (5.6) && 31.8 (5.8) && $<0.001$\\
\textbf{LDL cholesterol}, mmol/L &&& 2.3 (0.8) && 2.3 (0.8) && 2.3 (0.8) && $<0.001$\\
\textbf{Serum creatinine}, mmol/L &&& 81 (19) && 79 (18) && 82 (19) && $<0.001$\\
\textbf{Systolic BP} &&& 138 (10) && 137 (10) && 137 (10) && $<0.001$\\
\textbf{Diastolic BP} &&& 81 (6) && 81 (6) && 80 (6) && $<0.001$\\
\textbf{Triglycerides}, mmol/L &&& 2.3 (1.3) && 2.2 (1.2) && 2.2 (1.2) && $0.001$\\
\textbf{Comorbidity}, $n$ (\%) &&& && && \\
\ \ \ \ CVD &&& 837 (8.0) && 202 (6.4) && 113 (6.2) && $0.002$\\
\ \ \ \ CHF &&& 342 (3.3) && 90 (2.8) && 44 (2.4) && $0.107$\\
\ \ \ \ Stroke &&& 436 (4.2) && 100 (3.1) && 64 (3.5) && $0.025$\\
\ \ \ \ AF &&& 563 (5.4) && 147 (4.6) && 92 (5.1) && $0.244$\\
\ \ \ \ Cancer &&& 712 (6.8) && 173 (5.4) && 111 (6.1) && $0.021$\\
\ \ \ \ Arrhythmia &&& 626 (6.0) && 170 (5.3) && 97 (5.3) && $0.295$\\
\ \ \ \ COPD &&& 622 (5.9) && 133 (4.2) && 93 (5.1) && $0.001$\\ 
\ \ \ \ Hypertension &&& 1535 (14.6) && 349 (11.0) && 270 (14.9) && $<0.001$\\
\ \ \ \ Neuropathy &&& 573 (5.5) && 162 (5.1) && 118 (6.5) && $0.106$\\
\ \ \ \ Retinopathy &&& 2671 (25.5) && 712 (22.4) && 518 (28.5) && $<0.001$\\
\ \ \ \ Renal disease &&& 86 (0.8) && 18 (0.6) && 9 (0.5) && $0.154$\\
\ \ \ \ Falls &&& 799 (7.6) && 199 (6.3) && 170 (9.4) && $<0.001$\\
\textbf{Concurrent medication}, $n$ (\%) &&& && && \\
\ \ \ \ ACE/ARB &&& 6466 (61.7) && 1977 (62.2) && 1124 (61.9) && $0.871$\\
\ \ \ \ $\beta$-blockers &&& 214 (2.0) && 67 (2.1) && 27 (1.5) && $0.256$\\
\ \ \ \ CCBs &&& 2028 (19.3) && 637 (19.7) && 325 (17.9) && $0.263$\\
\ \ \ \ Statins &&& 8674 (82.7) && 2681 (84.3) && 1544 (85.0) && $0.014$\\
\ \ \ \ Antiarrhythmics &&& 135 (1.3) && 28 (0.9) && 17 (0.9) && $0.109$\\
\ \ \ \ Antipsychotics &&& 40 (0.4) && 9 (0.3) && 3 (0.2) && $0.287$\\
\ \ \ \ Loop diuretics &&& 935 (8.9) && 269 (8.5) && 100 (5.5) && $<0.001$\\
\ \ \ \ Nitrates &&& 389 (3.7) && 120 (3.8) && 53 (2.9) && $0.221$\\
\ \ \ \ Aspirin &&& 4347 (31.5) && 1211 (38.1) && 806 (44.4) && $<0.001$\\
\bottomrule
\multicolumn{10}{l}{$^{*}$Descriptive statistics are presented as the median (IQR) for continuous variables, unless otherwise indicated.}
\end{tabular}
\label{baseline.characteristics.support}
\end{table}

\subsection{Results}

\begin{figure}
\centering
\includegraphics[scale=0.45]{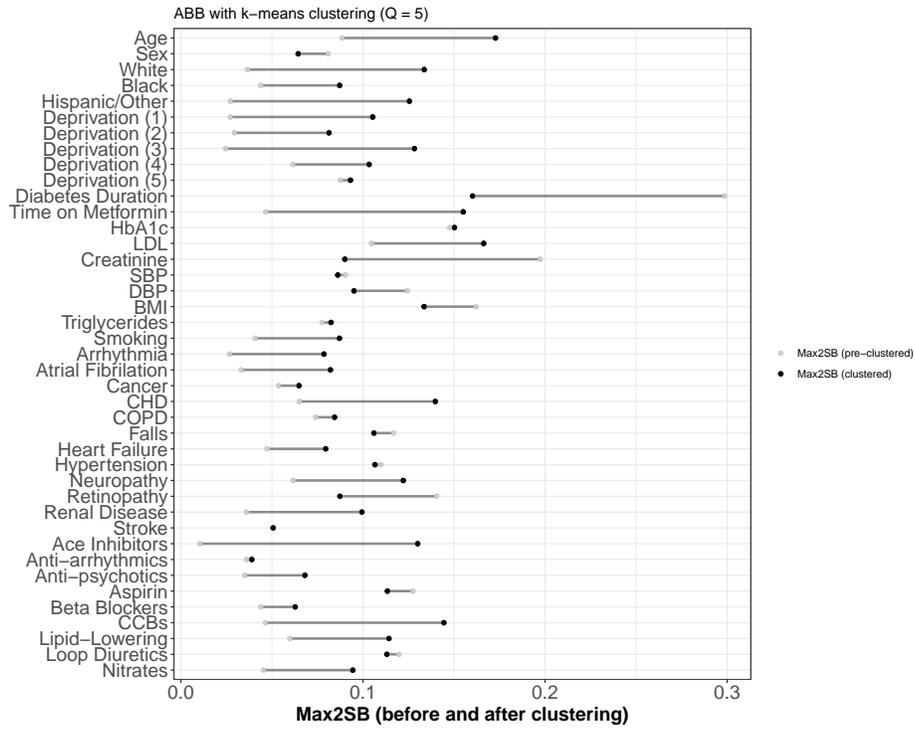}
\caption{$Max2SB_{p}$ before and after ABB with $k$-means clustering ($Q=5$)}
\label{diag_kmeans5}
\end{figure}

\begin{figure}
\centering
\includegraphics[scale=0.45]{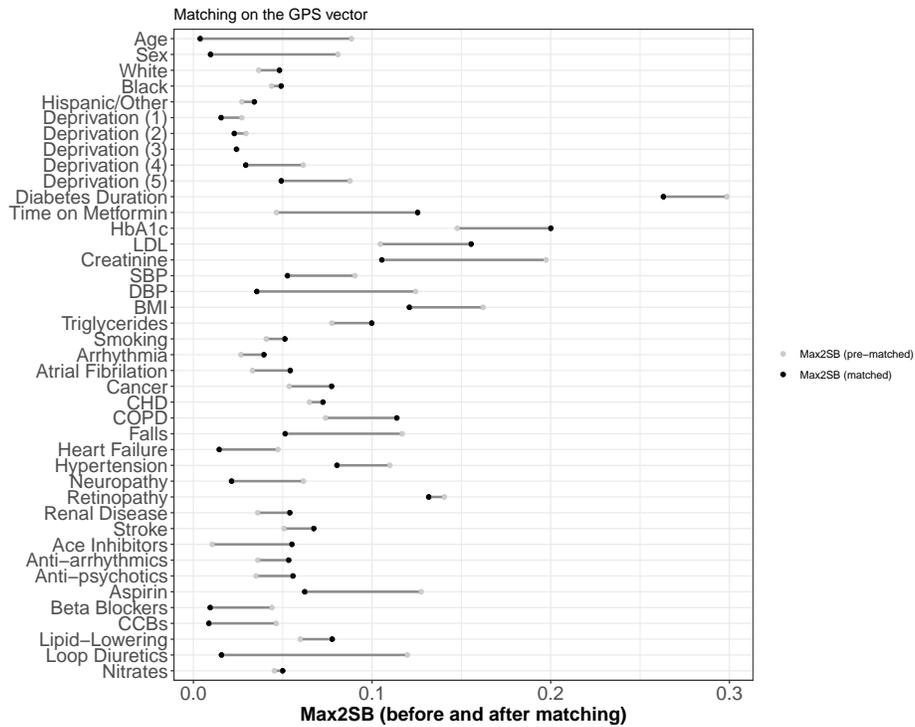}
\caption{$Max2SB_{p}$ before and after LGPSMnc matching}
\label{diag_matched}
\end{figure}

Covariate balance diagnostics in the form of $Max2SB_{p}$ in the initial cohort and the cohort after matching or clustering are given in Figures~\ref{diag_kmeans5} and \ref{diag_matched}. Reduction in initial bias was observed for most of the covariates using either clustering or matching. Except for diabetes duration with the matching algorithm, all of the $Max2SB_{p}$ were smaller than 0.2 for both clustering and matching algorithms. Compared to matching, ABB with $k$-means clustering with $Q=5$ yielded a larger reduction overall in covariate bias between treatment groups ($MaxMax2SB=0.1727$ for ABB with $Q=5$, $MaxMax2SB=0.2631$ for LGPSMnc). 

We set metformin plus gliclazide as the reference level. The estimated causal effects in the three-year risk of MACE or ACM using ABB with $k$-means clustering, matching, and IPW are depicted in Table~\ref{risk.table}. The baseline rates of MACE and ACM when using ABB for each of the three second-line treatment groups is given in Table~\ref{baseline.table} in the Appendix. Pairwise comparisons between gliclazide and pioglitazone yielded uniformly positive and statistically significant differences in three-year risk of MACE, with the effect size ranging from a 3.2\% (IPW) to a 4.6\% (ABB with $Q=1$) higher risk of MACE for gliclazide second-line users. Pairwise comparisons between sitagliptin and pioglitazone users also yielded uniformly positive and statistically significant differences in three-year risk of MACE. Pairwise comparisons between gliclazide and sitagliptin yielded uniformly positive and statistically significant differences in three-year risk of ACM, with the effect size ranging from 0.9\% (matching and IPW) to 1.6\% (ABB with $Q=1$) higher risk of ACM for gliclazide second-line users. Pairwise comparisons between gliclazide and pioglitazone  and sitagliptin and pioglitazone yielded uniformly small and non-statistically significant differences in three-year risk of ACM. Similar trends occur when using the three-year risk ratio with ABB, shown in Table~\ref{relative.risk.table} in the Appendix.

Generally, ABB with $Q=1$ reports larger differences than the other methods. The most notable difference in estimated three-year risk differences is observed when comparing gliclazide and sitagliptin  risk of MACE. ABB with $Q=1$ yielded a positive, statistically significant difference in three-year risk of MACE, with an effect size of 1.6\%. However, when using ABB with $Q\in\{3,5,7\}$, matching or IPW, the effect sizes are not statistically significant. These results are similar to those found in the simulations, where ABB with $Q=1$ results in be biased estimates and coverage rates that are below nominal when estimating a null ATT.  When comparing ABB with $Q \in \{5,7\}$ to matching and IPW, only minor differences are observed for the point and interval estimates of ACM. However, larger differences in point estimates are observed when comparing MACE for gliclazie and sitagliptin. Matching and IPW point estimates indicate that gliclazide has a lower probability of MACE compared sitagliptin and ABB with $Q\in\{5,7\}$ indicate a higher probability. The interval estimates of all these methods overlapped and included 0. 

\begin{table}[!h]
\centering 
\caption{Estimated 3-year risk difference (expressed as \%) for MACE and ACM (standard errors in parentheses) among gliclazide second-line users}
\begin{tabular}{l l l l l }
\toprule
Outcome & Method && Gliclazide vs. Sitagliptin & Gliclazide vs. Pioglitazone \\
\midrule
\underline{MACE} & \underline{ABB} && & \\
& $Q=1$ && 1.6 (0.8)$^{*}$ & 4.6 (0.8)$^{*}$ \\
& $Q=3$ && 0.1 (0.8) & 4.4 (0.9)$^{*}$ \\
& $Q=5$ && 0.1 (0.9) & 3.7 (0.9)$^{*}$ \\
& $Q=7$ && 0.1 (0.9) & 3.5 (0.9)$^{*}$ \\
& \underline{Matching} && -0.1 (0.8) & 3.6 (1.0)$^{*}$ \\
& \underline{IPW} && -0.4 (0.8) & 3.2 (0.9)$^{*}$ \\
\midrule\midrule
\underline{ACM} & \underline{ABB} && & \\
& $Q=1$ && 1.6 (0.4)$^{*}$ & 0.7 (0.5) \\
& $Q=3$ && 1.2 (0.4)$^{*}$ & 0.4 (0.6) \\
& $Q=5$ && 1.2 (0.5)$^{*}$ & 0.3 (0.6) \\
& $Q=7$ && 1.1 (0.4)$^{*}$ & 0.5 (0.5) \\
& \underline{Matching} && 0.9 (0.4)$^{*}$ & 0.5 (0.6) \\
& \underline{IPW} && 0.9 (0.4)$^{*}$ & 0.3 (0.5) \\
\bottomrule
\multicolumn{5}{l}{$^{*}$Significant at the 0.05 level}
\end{tabular}
\label{risk.table}
\end{table}

\subsection{Sensitivity analysis}

All of the proposed methods rely on the strong ignorability assumption which is untestable with observed data. To examine the validity of this assumption with ABB, we propose an interpretable sensitivity analysis. 

We examine the sensitivity of the $\tau^{G}_{G,S}$ and $\tau^{G}_{G,P}$ to an unmeasured covariate by introducing a continuous variable $\xi$ that is independent from the observed covariates. We assume that $\xi_i$ is Normally distributed with mean $\delta Y_{i}^{obs} + \phi  T_{i,S} + \phi T_{i,P}$ and a variance of 1. In this model, $\phi$ represents the initial bias in $\xi_i$ when comparing gliclazide to sitagliptin or gliclazide to pioglitazone, and $\exp(\delta)$ represents the odds ratio of $Y_i^{obs}\in\{0,1\}$ given an increase of one unit in $\xi_{i}$.

To provide intuition to this sensitivity analysis, it is useful to consider two possible scenarios. The first scenario is when the signs of $\delta$ and $\phi$ are the same. Under this scenario, the distribution of $\xi$ with $Y_{i}^{obs}=1$ among gliclazide users is more similar to the distribution of $\xi$ with $Y_{i}^{obs}=0$ among pioglitazone or sitagliptin users. Thus, these units would be more similar when $\xi$ is included in the generalized propensity score estimation, which will yield a larger estimated ATT. In the second scenario, the signs of $\delta$ and $\phi$ are opposite. As a consequence, the distribution of $\xi$ with $Y_{i}^{obs}=0$ among gliclazide users is more similar to the distribution of $\xi$ with $Y_{i}^{obs}=1$ among pioglitazone or sitagliptin users. This scenario will yield a smaller estimated ATT when $\xi$ is included in the generalized propensity score estimation.

\begin{figure}[!h]
\centering
\includegraphics[scale=0.4]{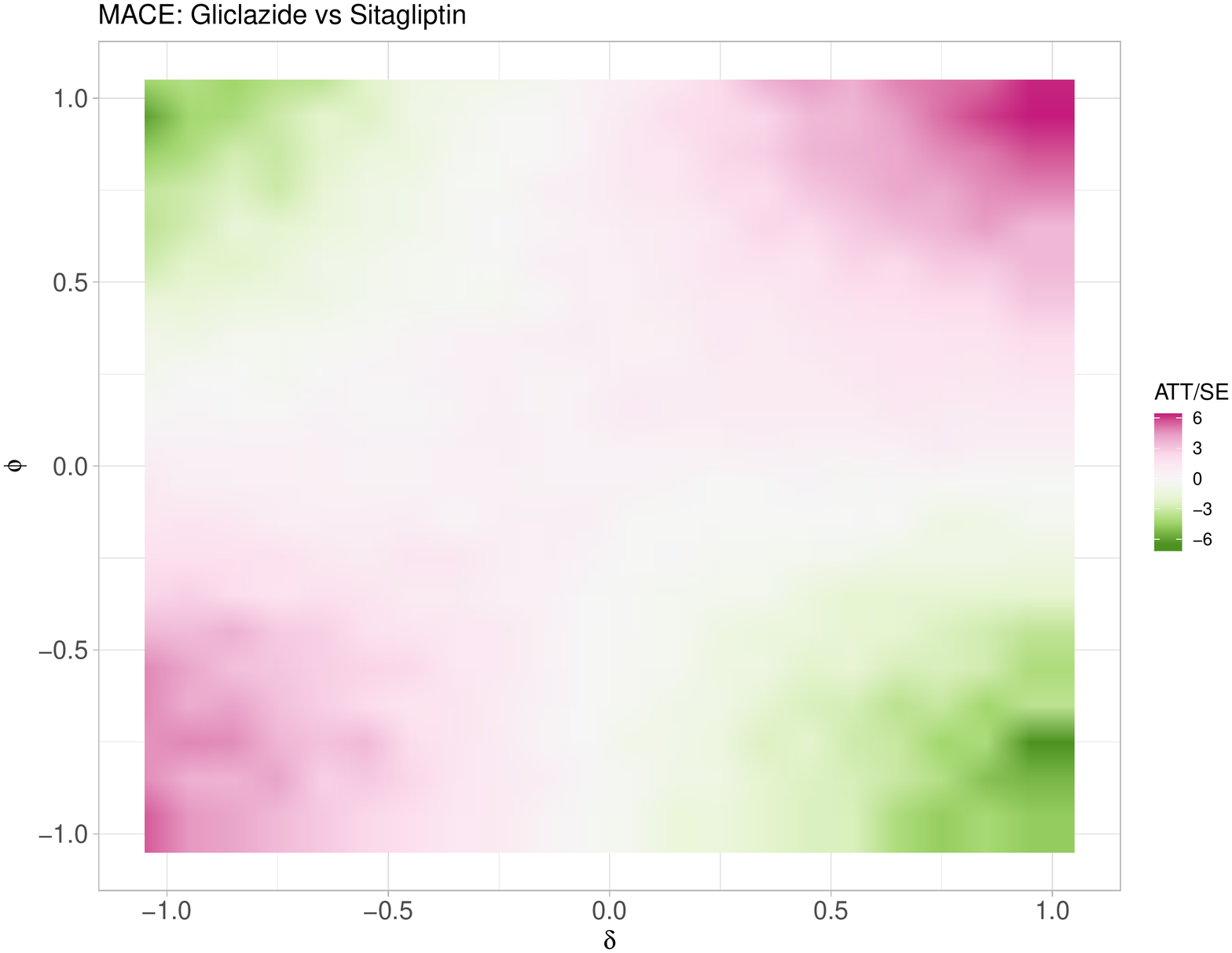}
\includegraphics[scale=0.4]{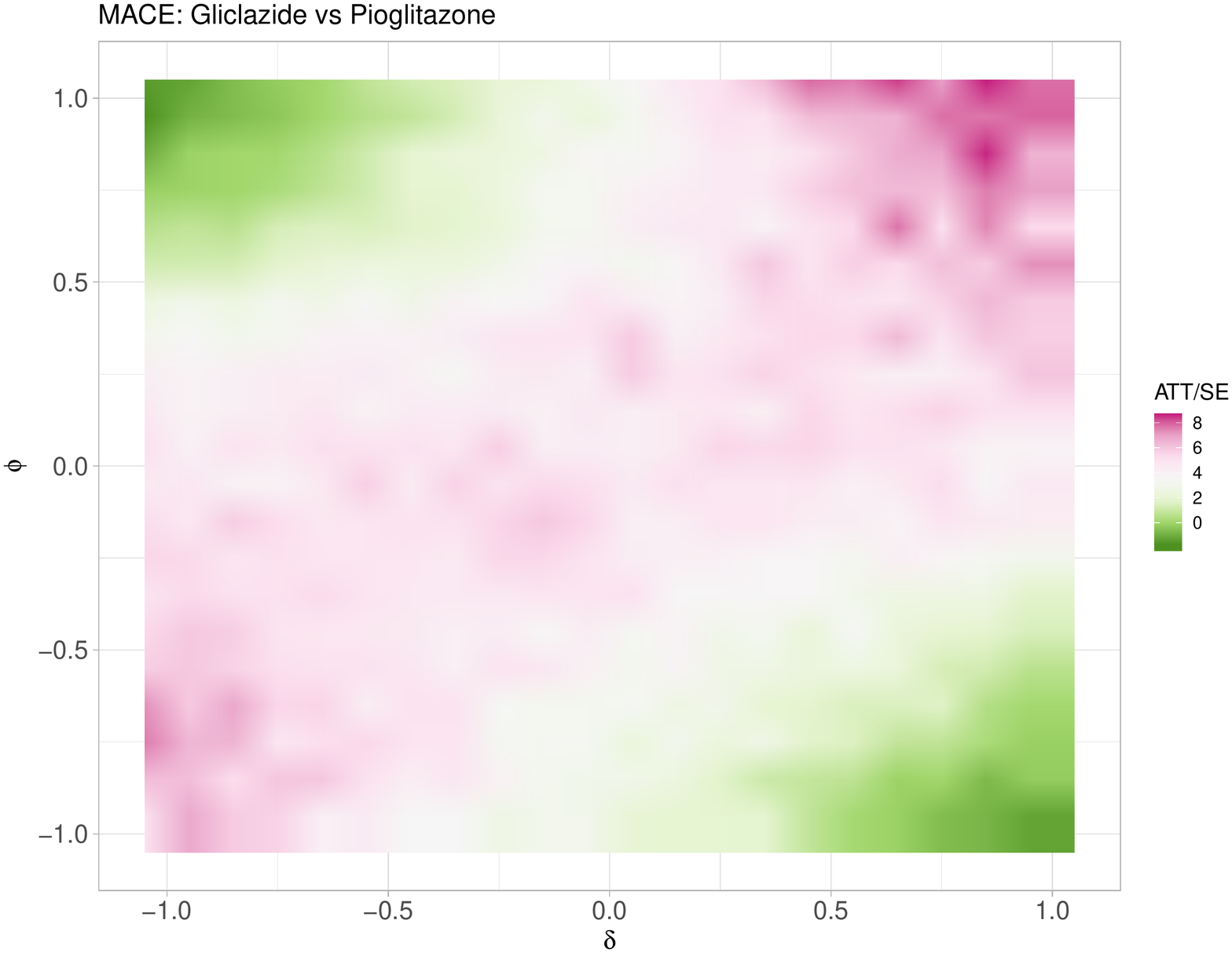}
\caption{Standardized effect for varying $\delta$ and $\phi$, 3-year MACE risk}
\label{mace.sensitivity}
\end{figure}

We estimated $\hat{\tau}^{G}_{G,S}$ and $\hat{\tau}^{G}_{G,P}$ for MACE using the observed data and $\xi$. The GPS estimation included $\xi$ as another covariate in the multinomial logistic regression model, and used ABB with $k$-means clustering that assumed $Q=5$. Figure~\ref{mace.sensitivity} displays the standardized effects, $\hat{\tau}^{G}_{G,S}/SE(\hat{\tau}^{G}_{G,S})$ and $\hat{\tau}^{G}_{G,P}/SE(\hat{\tau}^{G}_{G,P})$ for MACE, for $\phi \in (-1,1)$ and for $\delta \in (-1,1)$. 

Based on Table~\ref{risk.table}, the standardized difference in MACE risks between gliclazide and sitagliptin is close to zero before introducing $\xi$  (e.g. $\phi=0$ and $\delta=0$). When incorporating $\xi$ into the estimation procedure, the standardized effects are most sensitive to configurations with $|\delta|\geq0.5$ and $|\phi|\geq0.5$. The absolute value of the standardized effect was greater than 2 for these configurations. This implies that in order to find a statistically significant effect, the unobserved covariate should have a standardized initial bias greater than 0.5 or an odds ratio that is greater than $\exp(0.5)$. The standardized difference in MACE risks between gliclazide and pioglitazone is equal to 3.99 before introducing $\xi$. When incorporating $\xi$ in these comparisons, the standardized effects are most sensitive to $\{0.5\leq\delta<1\}\cap\{-1<\phi\leq-0.5\}$ and $\{-1<\delta\leq-0.5\}\cap\{0.5\leq\phi<1\}$, where the standardized effect approached 0 for these configurations. None of the observed covariates had initial bias that is greater than 0.3. In addition, an odds ratio of 1.6 is similar to a moderate Cohen's $d$ \cite{chen-10}. Thus, we can conclude that the estimates are relatively robust to the strong ignorability assumption.

\section{Discussion}

Many problems in public health involve comparing three or more interventions. This paper proposes an approximate Bayesian bootstrap (ABB) procedure that uses $k$-means clustering on the generalized propensity score (GPS) to estimate causal effects of multiple nominal treatments. The $k$-means clustering component ensures that donors have relatively similar covariate values, and the ABB component imputes the unobserved potential outcomes. This procedure can be used with binary or categorical outcomes, and it provides a statistically valid inference procedure for any estimand. We also describe a sensitivity analysis to the strong ignorability assumption. The sensitivity analysis can be implemented within the proposed ABB procedure and it is interpretable using graphical visualization.

Using simulations, we compared the performance of the newly proposed procedures to matching and IPW estimators. For these simulation configurations we observed that ABB with $k$-means clustering is generally accurate and precise when using 5 or 7 clusters compared to weighting and matching methods. 

The simulation study used $Z=3$ treatment groups and the logit GPS vector as a distance measure for $k$-means clustering and matching. Scotina and Gutman \cite{scotina-19_MA} found that this distance measure can perform poorly  in terms of reducing covariate bias between treatment groups, when used in matching algorithms with $Z>5$. An area for future work would be to identify the operating characteristics for ABB multiple imputation with $k$-means clustering when there is a large number of treatments. ABB with $k$-means clustering only imputes the missing potential outcomes based on the GPS. A potential extension would be to impute missing potential outcomes based on a model that includes the generalized propensity score and covariates that predict the outcome. This type of modeling was shown to yield more precise and accurate estimates with two treatments \cite{gutman-15, gutman-13}. 

For the nearly 20 million people affected with type 2 diabetes in the United States, acute myocardial infarction is a very serious and potentially fatal complication of the disease, the choices of treatments are highly complex \cite{ada-14}, and the effects of various antihyperglycemic treatments on macrovascular events is uncertain \cite{hiatt-13, smith-13}. The analysis presented here is a novel approach to comparing the risks of major adverse cardiovascular events (MACE) and all-cause mortality (ACM) of multiple second-line antidiabetic medications using observational data \cite{seong-15, ekstrom-16, obrien-18}. We show that, among patients who initiate second-line treatment after metformin monotherapy, there is a significantly higher three-year risk of MACE for users of gliclazide and sitagliptin compared to users of pioglitazone. Further, there is a significantly higher three-year risk of ACM for gliclazide second-line users compared to sitagliptin second-line users. It will be of interest to explore the use of this methodology to investigate the effects of these antihyperglycemic agents and other members of the sulfonylurea and DPP-4i drug classes in other large clinical databases. As more real-world clinical data with new antihyperglycemic agents that have shown evidence for cardiovascular benefit, such as members of the sodium glucose co-transporter 2 inhibitor class \cite{zinman-15, neal-17}, become available, it will be of interest to determine whether the observations from prospective, randomized clinical trials with these agents are confirmed. 

In conclusion, we provide a general method for estimating causal effects using the generalized propensity score  that can be applied to binary or categorical outcomes and any estimand of interest. Based on our simulations, ABB with $k$-means clustering using $K\in\{5,7\}$ clusters generally yields valid, accurate, and precise estimates of average treatment effects.

\section*{Acknowledgment}

This research was supported through a Patient-Centered Outcomes Research Institute (PCORI) Award ME-1403-12104. Disclaimer: All statements in this report, including its findings and conclusions, are solely those of the authors and do not necessarily represent the views
of the PCORI, its Board of Governors or Methodology Committee.

\newpage

\bibliographystyle{SageV}
\bibliography{references}

\newpage

\section{Appendix}

\subsection{Cohort information}

\begin{figure}[!h]
\centering
  \begin{tikzpicture}[auto,
    block_center/.style ={rectangle, draw=black, thick, fill=white,
      text width=20em, text centered,
      minimum height=4em},
    block_left/.style ={rectangle, draw=black, thick, fill=white,
      text width=12em, text ragged, minimum height=4em, inner sep=6pt},
     block_left2/.style ={rectangle, draw=black, thick, fill=white,
      text width=20em, text ragged, minimum height=4em, inner sep=6pt},
    block_noborder/.style ={rectangle, draw=none, thick, fill=none,
      text width=18em, text centered, minimum height=1em},
    block_assign/.style ={rectangle, draw=black, thick, fill=white,
      text width=18em, text ragged, minimum height=3em, inner sep=6pt},
    block_lost/.style ={rectangle, draw=black, thick, fill=white,
      text width=16em, text ragged, minimum height=3em, inner sep=6pt},
      line/.style ={draw, thick, -latex', shorten >=0pt}]
    \matrix [column sep=5mm,row sep=3mm] {
      \node [block_center] (referred) {Patients treated with metformin between January 2004 and December 2015 ($n=103690$)};
      & \node [block_left] (excluded1) {Excluded ($n=44215$): \\
      \h Patients on metformin \\
      \h monotherapy only}; \\
      \node [block_center] (assessment) {Patients treated with metformin monotherapy for at least 60 days ($n=34959$)}; 
      & \node [block_left] (excluded2) {Excluded ($n=4274$): \\
        \h Patients who changed \\
        \h from metformin mono-\\
        \h therapy to monotherapy \\
        \h of a different agent}; \\
      \node [block_center] (second) {Patients who started continuous treatment of a second non-insulin antihyperglycemic agent and who maintain continuous treatment of metformin ($n=25924$)};
      & \node [block_left] (excluded3) {Excluded ($n=981$): \\
      \h Patients who used SGLT-\\
      \h 2i or GLP-1RA as second-\\
      \h line agent}; \\
      \node [block_center] (remove) {Patients age 18-85, BMI 18.5-50 kg/m$^{2}$, serum creatinine 20-250 $\mu$mol/l, and second-line treatment initiation 2007 or later ($n=21976$)};\\
      \node [block_center] (excludeagents) {Patients who used gliclazide, sitagliptin, or pioglitazone as second-line agent ($n=18543)$ };\\
      & \node [block_left] (excluded4) {Excluded ($n=3059$): \\
        \h Patients outside the \\
        \h region of\\
        \h common support}; \\
      \node [block_left2] (treatments) {Treatment regimens:\\
      \h 1) Metformin + SU ($n=10487$)\\
      \h 2) Metformin + DPP-4 ($n=3180$)\\
      \h 3) Metformin + TZD ($n=1817$)};\\
    };
    \begin{scope}[every path/.style=line]
      \path (referred)   -- (excluded1);
      \path (referred)   -- (assessment);
      \path (assessment) -- (excluded2);
      \path (assessment) -- (second);
      \path (second) -- (excluded3);
      \path (second) -- (remove);
      \path (remove) -- (excludeagents); 
      \path (excludeagents) -- (excluded4);
      \path (excludeagents) -- (treatments);
    \end{scope}
  \end{tikzpicture}
\caption{Description of cohort selection}
\label{flowchart}
\end{figure}
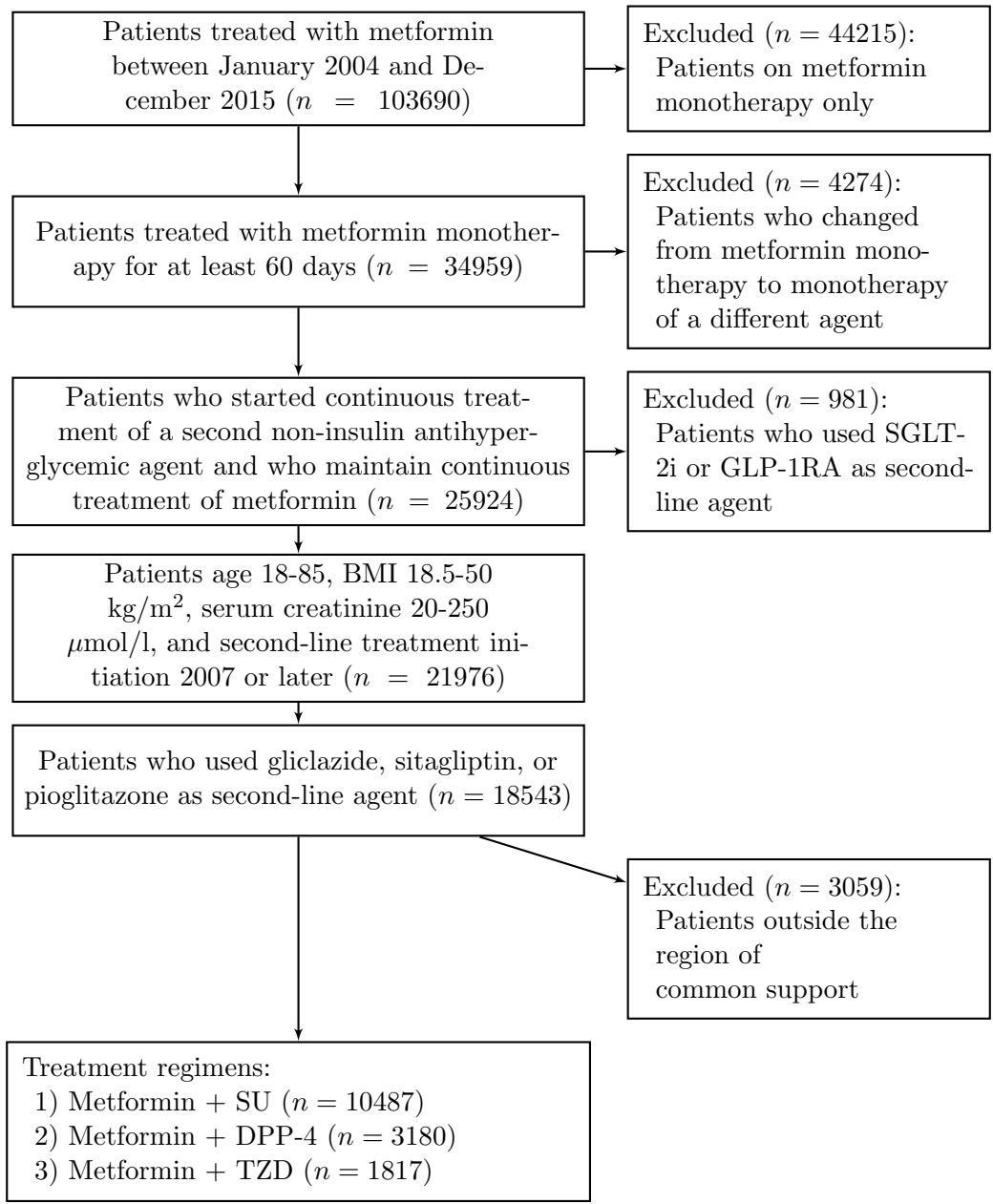 

\newpage

\subsection{ICD-10 Codes for Outcome Events}

\begin{itemize}
	\item I20-I25: Ischemic heart diseases
	\item I70.2: Atherosclerosis of native arteries of the extremities
	\item I73.1: Thromboangiitis obliterans
	\item I73.9: Peripheral vascular disease (unspecified)
	\item E10.5: Type 1 diabetes mellitus with circulatory complications
	\item E11.5: Type 2 diabetes mellitus with circulatory complications
	\item I50: Heart failure
	\item I61-64: Nontraumatic intracerebral hemorrhage, other and unspecified nontraumatic intracranial hemorrhage, cerebral infarction, stroke (unspecified)
\end{itemize}

\newpage

\begin{sidewaystable}[!ht]
\small
\centering
\caption{Simulated coverage, bias, and standard deviation of bias of log odds ratio estimators across levels of $b$, when treatment effect is null}
\begin{tabular}{l l c c c c c c c}
\toprule
& & \multicolumn{7}{c}{Treatment Group Comparison} \\
\cline{3-9}\\
& & \multicolumn{3}{c}{1 vs. 2} & & \multicolumn{3}{c}{1 vs. 3}\\
\cline{3-5}  \cline{7-9} \\
$b$ & Method & Coverage & Mean absolute bias (SD) & Std. Error && Coverage & Mean absolute bias (SD) & Std. Error \\
0.00 & \underline{ABB} & & & & & & & \\
& $Q=1$ & 1.00  & 0.0592 (0.0494) & 0.1084 && 1.00 & 0.0549 (0.0385) & 0.1048 \\
& $Q=3$ & 1.00 & 0.0441 (0.0328) & 0.1090 && 1.00 & 0.0380 (0.0292) & 0.1046 \\
& $Q=5$ & 1.00 & 0.0471 (0.0365) & 0.1089 && 1.00 & 0.0417 (0.0291) & 0.1038 \\
& $Q=7$ & 1.00 & 0.0390 (0.0306) & 0.1059 && 1.00 & 0.0330 (0.0286) & 0.1038 \\ 
\midrule
0.25 & \underline{ABB} & & & & & & & \\
& $Q=1$ & 0.99 & 0.0971 (0.0610) & 0.1137 && 0.81 & 0.1732 (0.0679) & 0.1140 \\
& $Q=3$ & 1.00 & 0.0763 (0.0529) & 0.1171 && 0.99 & 0.1007 (0.0501) & 0.1101 \\
& $Q=5$ & 1.00 & 0.0454 (0.0332) & 0.1156 && 1.00 & 0.0808 (0.0422) & 0.1107 \\ 
& $Q=7$ & 1.00 & 0.0655 (0.0468) & 0.1173 && 1.00 & 0.0461 (0.0377) & 0.1097 \\
\midrule
0.50 & \underline{ABB} & & & & & & & \\
& $Q=1$ & 0.75 & 0.2042 (0.0751) & 0.1279 && 0.04 & 0.3650 (0.0653) & 0.1398 \\
& $Q=3$ & 1.00 & 0.1204 (0.0692) & 0.1385 && 0.82 & 0.1975 (0.0676) & 0.1299 \\
& $Q=5$ & 1.00 & 0.0704 (0.0549) & 0.1363 && 1.00 & 0.1001 (0.0567) & 0.1380 \\
& $Q=7$ & 1.00 & 0.0774 (0.0598) & 0.1408 && 1.00 & 0.0735 (0.0548) & 0.1303 \\
\midrule
0.75 & \underline{ABB} & & & & & & & \\
& $Q=1$ & 0.65 & 0.2539 (0.0843) & 0.1464 && 0.01 & 0.5010 (0.0729) & 0.1743 \\
& $Q=3$ & 0.98 & 0.1339 (0.0895) & 0.1678 && 0.85 & 0.2007 (0.1032) & 0.1503 \\
& $Q=5$ & 0.96 & 0.1151 (0.0916) & 0.1820 && 0.99 & 0.0853 (0.0594) & 0.1579 \\
& $Q=7$ & 0.90 & 0.1476 (0.1001) & 0.1747 && 0.98 & 0.1079 (0.0711) & 0.1607 \\
\midrule
1.00 & \underline{ABB} & & & & & & & \\
& $Q=1$ & 0.39 & 0.3863 (0.0934) & 0.1848 && 0.00 & 0.7788 (0.0913) & 0.2515 \\
& $Q=3$ & 0.92 & 0.1584 (0.1219) & 0.2128 && 0.99 & 0.1318 (0.0980) & 0.1911 \\
& $Q=5$ & 0.87 & 0.2202 (0.1671) & 0.2477 && 0.89 & 0.1689 (0.1324) & 0.2124 \\
& $Q=7$ & 0.85 & 0.2514 (0.1753) & 0.2644 && 0.74 & 0.2650 (0.1654) & 0.2289 \\
\bottomrule
\end{tabular}
\label{sim.OR.null}
\end{sidewaystable}

\newpage

\begin{sidewaystable}[!ht]
\small
\centering
\caption{Simulated coverage, bias, and standard deviation of bias of log odds ratio estimators across levels of $b$, when treatment effect is non-null}
\begin{tabular}{l l c c c c c c c}
\toprule
& & \multicolumn{7}{c}{Treatment Group Comparison} \\
\cline{3-9}\\
& & \multicolumn{3}{c}{1 vs. 2} & & \multicolumn{3}{c}{1 vs. 3}\\
\cline{3-5}  \cline{7-9} \\
$b$ & Method & Coverage & Mean absolute bias (SD) & Std. Error && Coverage & Mean absolute bias (SD) & Std. Error \\
0.00 & \underline{ABB} & & & & & & & \\
& $Q=1$ & 1.00 & 0.0485 (0.0440) & 0.1026 && 1.00 & 0.0523 (0.0368) & 0.0988 \\
& $Q=3$ & 1.00 & 0.0455 (0.0324) & 0.1013 && 1.00 & 0.0370 (0.0306) & 0.0983 \\
& $Q=5$ & 1.00 & 0.0415 (0.0311) & 0.1001 && 1.00 & 0.0372 (0.0282) & 0.0973 \\
& $Q=7$ & 1.00 & 0.0393 (0.0302) & 0.1015 && 1.00 & 0.0315 (0.0272) & 0.0989 \\ 
\midrule
0.25 & \underline{ABB} & & & & & & & \\
& $Q=1$ & 1.00 & 0.0640 (0.0488) & 0.1027 && 0.820 & 0.1491 (0.0646) & 0.1042 \\
& $Q=3$ & 1.00 & 0.0703 (0.0429) & 0.1086 && 1.00 & 0.0808 (0.0453) & 0.1021 \\
& $Q=5$ & 1.00 & 0.0589 (0.0417) & 0.1067 && 0.99 & 0.0723 (0.0497) & 0.1021 \\ 
& $Q=7$ & 1.00 & 0.0466 (0.0348) & 0.1063 && 1.00 & 0.0548 (0.0350) & 0.1007 \\
\midrule
0.50 & \underline{ABB} & & & & & & & \\
& $Q=1$ & 0.66 & 0.1914 (0.0793)& 0.1131 && 0.04 & 0.3446 (0.0684) & 0.1216 \\
& $Q=3$ & 0.91 & 0.1502 (0.0762) & 0.1254 && 0.81 & 0.1864 (0.0586) & 0.1168 \\
& $Q=5$ & 0.99 & 0.1191 (0.0594) & 0.1248 && 0.99 & 0.1073 (0.0560) & 0.1158 \\
& $Q=7$ & 0.99 & 0.0652 (0.0534) & 0.1237 && 0.99 & 0.0732 (0.0499) & 0.1155 \\
\midrule
0.75 & \underline{ABB} & & & & & & & \\
& $Q=1$ & 0.45 & 0.2602 (0.0868) & 0.1286 && 0.00 & 0.5351 (0.0743) & 0.1497 \\
& $Q=3$ & 0.97 & 0.1340 (0.0875) & 0.1425 && 0.72 & 0.2268 (0.0873) & 0.1371 \\
& $Q=5$ &  0.99 & 0.1120 (0.0805) & 0.1509 && 1.00 & 0.0947 (0.0692) & 0.1384 \\
& $Q=7$ & 0.89 & 0.1279 (0.1013) & 0.1588 && 0.93 & 0.1071 (0.0821) & 0.1402 \\
\midrule
1.00 & \underline{ABB} & & & & & & & \\
& $Q=1$ & 0.21 & 0.3733 (0.0889) & 0.1562 && 0.00 & 0.6758 (0.0805) & 0.1936 \\
& $Q=3$ & 0.93 & 0.1643 (0.1053) & 0.1809 && 1.00 & 0.1277 (0.0881) & 0.1620 \\
& $Q=5$ & 0.78 & 0.2280 (0.1453) & 0.2050 && 0.87 & 0.1679 (0.1206) & 0.1813 \\
& $Q=7$ & 0.72 & 0.2665 (0.1963) & 0.2110 && 0.94 & 0.1822 (0.1267) & 0.2140 \\
\bottomrule
\end{tabular}
\label{sim.OR.nonnull}
\end{sidewaystable}

\newpage

\begin{sidewaystable}[!ht]
\small
\centering
\caption{Simulated coverage, bias, and standard deviation of bias of log risk ratio estimators across levels of $b$, when treatment effect is null}
\begin{tabular}{l l c c c c c c c}
\toprule
& & \multicolumn{7}{c}{Treatment Group Comparison} \\
\cline{3-9}\\
& & \multicolumn{3}{c}{1 vs. 2} & & \multicolumn{3}{c}{1 vs. 3}\\
\cline{3-5}  \cline{7-9} \\
$b$ & Method & Coverage & Mean absolute bias (SD) & Std. Error && Coverage & Mean absolute bias (SD) & Std. Error \\
0.00 & \underline{ABB} & & & & & & & \\
& $Q=1$ & 0.9900 & 0.0292 (0.0247) & 0.0538 && 1.00 & 0.0279 (0.0193) & 0.0525 \\
& $Q=3$ & 1.00 & 0.0219 (0.0164) & 0.0544 && 1.00 & 0.0192 (0.0148) & 0.0525 \\
& $Q=5$ & 1.00 & 0.0235 (0.0184) & 0.0542 && 1.00 & 0.0209 (0.0147) & 0.0520 \\
& $Q=7$ & 1.00 & 0.0199 (0.0159) & 0.0543 && 1.00 & 0.0168 (0.0146) & 0.0527 \\ 
\midrule
0.25 & \underline{ABB} & & & & & & & \\
& $Q=1$ & 0.98 & 0.0392 (0.0250) & 0.0457 && 0.76 & 0.0688 (0.0276) & 0.0430 \\
& $Q=3$ & 1.00 & 0.0305 (0.0217) & 0.0472 && 0.96 & 0.0407 (0.0209) & 0.0438 \\
& $Q=5$ & 1.00 & 0.0189 (0.0137) & 0.0484 && 1.00 & 0.0328 (0.0175) & 0.0449 \\ 
& $Q=7$ & 1.00 & 0.0268 (0.0191) & 0.0481 && 1.00 & 0.0189 (0.0158) & 0.0452 \\
\midrule
0.50 & \underline{ABB} & & & & & & & \\
& $Q=1$ & 0.63 & 0.0656 (0.0250) & 0.0385 && 0.01 & 0.1123 (0.0219) & 0.0356 \\
& $Q=3$ & 1.00 & 0.0390 (0.0227) & 0.0437 && 0.70 & 0.0639 (0.0227) & 0.0389 \\
& $Q=5$ & 1.00 & 0.0238 (0.0185) & 0.0462 && 1.00 & 0.0323 (0.0192) & 0.0434 \\
& $Q=7$ & 0.99 & 0.0257 (0.0193) & 0.0474 && 1.00 & 0.0238 (0.0176) & 0.0425 \\
\midrule
0.75 & \underline{ABB} & & & & & & & \\
& $Q=1$ & 0.46 & 0.0622 (0.0216) & 0.0316 && 0.00 & 0.1125 (0.0184) & 0.0290 \\
& $Q=3$ & 0.95 & 0.0336 (0.0229) & 0.0408 && 0.72 & 0.0505 (0.0261) & 0.0351 \\
& $Q=5$ & 0.98 & 0.0365 (0.0273) & 0.0526 && 0.99 & 0.0231 (0.0160) & 0.0431 \\
& $Q=7$ & 0.94 & 0.0464 (0.0311) & 0.0531 && 0.99 & 0.0317 (0.0221) & 0.0468 \\
\midrule
1.00 & \underline{ABB} & & & & & & & \\
& $Q=1$ & 0.16 & 0.0734 (0.0188) & 0.0272 && 0.00 & 0.1303 (0.0175) & 0.0241 \\
& $Q=3$ & 0.94 & 0.0364 (0.0299) & 0.0490 && 0.97 & 0.0266 (0.0194) & 0.0374 \\
& $Q=5$ & 0.93 & 0.0609 (0.0474) & 0.0669 && 0.95 & 0.0404 (0.0351) & 0.0510 \\
& $Q=7$ & 0.89 & 0.0653 (0.0508) & 0.0661 && 0.87 & 0.0744 (0.0490) & 0.0662 \\
\bottomrule
\end{tabular}
\label{sim.RR.null}
\end{sidewaystable}

\newpage

\begin{sidewaystable}[!ht]
\small
\centering
\caption{Simulated coverage, bias, and standard deviation of bias of log risk ratio estimators across levels of $b$, when treatment effect is non-null}
\begin{tabular}{l l c c c c c c c}
\toprule
& & \multicolumn{7}{c}{Treatment Group Comparison} \\
\cline{3-9}\\
& & \multicolumn{3}{c}{1 vs. 2} & & \multicolumn{3}{c}{1 vs. 3}\\
\cline{3-5}  \cline{7-9} \\
$b$ & Method & Coverage & Mean absolute bias (SD) & Std. Error && Coverage & Mean absolute bias (SD) & Std. Error \\
0.00 & \underline{ABB} & & & & & & & \\
& $Q=1$ & 1.00 & 0.0250 (0.0225) & 0.0557 && 1.00 & 0.0267 (0.0189) & 0.0536 \\
& $Q=3$ & 1.00 & 0.0237 (0.0167) & 0.0552 && 1.00 & 0.0187 (0.0157) & 0.0538 \\
& $Q=5$ & 1.00 & 0.0219 (0.0166) & 0.0552 && 1.00 & 0.0195 (0.0147) & 0.0533 \\
& $Q=7$ & 1.00 & 0.0202 (0.0157) & 0.0555 && 1.00 & 0.0159 (0.0138) & 0.0534 \\ 
\midrule
0.25 & \underline{ABB} & & & & & & & \\
& $Q=1$ & 1.00 & 0.0271 (0.0209) & 0.0472 && 0.80 & 0.0649 (0.0269) & 0.0442 \\
& $Q=3$ & 1.00 & 0.0306 (0.0187) & 0.0490 && 1.00 & 0.0357 (0.0196) & 0.0457 \\
& $Q=5$ & 1.00 & 0.0255 (0.0182) & 0.0484 && 0.99 & 0.0319 (0.0217) & 0.0464 \\ 
& $Q=7$ & 1.00 & 0.0201 (0.0150) & 0.0492 && 1.00 & 0.0240 (0.0153) & 0.0463 \\
\midrule
0.50 & \underline{ABB} & & & & & & & \\
& $Q=1$ & 0.58 & 0.0700 (0.0277) & 0.0391 && 0.01 & 0.1198 (0.0232) & 0.0370 \\
& $Q=3$ & 0.89 & 0.0541 (0.0267) & 0.0441 && 0.68 & 0.0676 (0.0204) & 0.0402 \\
& $Q=5$ & 0.98 & 0.0433 (0.0213) & 0.0466 && 0.98 & 0.0396 (0.0203) & 0.0434 \\
& $Q=7$ & 0.99 & 0.0243 (0.0201) & 0.0492 && 0.99 & 0.0255 (0.0176) & 0.0438 \\
\midrule
0.75 & \underline{ABB} & & & & & & & \\
& $Q=1$ & 0.38 & 0.0748 (0.0235) & 0.0335 && 0.00 & 0.1395 (0.0197) & 0.0298 \\
& $Q=3$ & 0.95 & 0.0379 (0.0241) & 0.0420 && 0.59 & 0.0641 (0.0236) & 0.0360 \\
& $Q=5$ &  0.99 & 0.0362 (0.0255) & 0.0526 && 0.99 & 0.0275 (0.0201) & 0.0449 \\
& $Q=7$ & 0.94 & 0.0419 (0.0354) & 0.0555 && 0.97 & 0.0345 (0.0271) & 0.0486 \\
\midrule
1.00 & \underline{ABB} & & & & & & & \\
& $Q=1$ & 0.10 & 0.0806 (0.0179) & 0.0280 && 0.00 & 0.1297 (0.0153) & 0.0245 \\
& $Q=3$ & 0.96 & 0.0410 (0.0280) & 0.0485 && 0.99 & 0.0283 (0.0187) & 0.0374 \\
& $Q=5$ & 0.96 & 0.0680 (0.0448) & 0.0665 && 0.94 & 0.0442 (0.0351) & 0.0540 \\
& $Q=7$ & 0.85 & 0.0840 (0.0680) & 0.0754 && 0.95 & 0.0469 (0.0321) & 0.0617 \\
\bottomrule
\end{tabular}
\label{sim.RR.nonnull}
\end{sidewaystable}

\newpage

\begin{sidewaystable}[!ht]
\small
\centering
\caption{Simulated coverage, bias, and standard deviation of bias of ATT estimators across levels of $b$, when outcome is ordinal}
\begin{tabular}{l l c c c c c c c}
\toprule
& & \multicolumn{7}{c}{Treatment Group Comparison} \\
\cline{3-9}\\
& & \multicolumn{3}{c}{1 vs. 2} & & \multicolumn{3}{c}{1 vs. 3}\\
\cline{3-5}  \cline{7-9} \\
$b$ & Method & Coverage & Mean absolute bias (SD) & Std. Error && Coverage & Mean absolute bias (SD) & Std. Error \\
0.00 & \underline{ABB} & & & & & & & \\
& $Q=1$ & 1.00 & 0.0451 (0.0322) & 0.0875 && 1.00 & 0.0447 (0.0334) & 0.0845 \\
& $Q=3$ & 1.00 & 0.0347 (0.0258) & 0.0861 && 1.00 & 0.0314 (0.0244) & 0.0830 \\
& $Q=5$ & 1.00 & 0.0362 (0.0229) & 0.0858 && 1.00 & 0.0357 (0.0280) & 0.0832 \\
& $Q=7$ & 1.00 & 0.0409 (0.0291) & 0.0860 && 1.00 & 0.0393 (0.0277) & 0.0831 \\ 
& \underline{Matching} & 0.98 & 0.0452 (0.0374) & 0.0745 && 0.99 & 0.0453 (0.0329) & 0.0698\\
& \underline{IPW} & 1.00 & 0.0327 (0.0245) & 0.0579 && 1.00 & 0.0314 (0.0231) & 0.0527\\
\midrule
0.25 & \underline{ABB} & & & & & & & \\
& $Q=1$ & 0.94 & 0.0703 (0.0410) & 0.0747 && 0.69 & 0.1155 (0.0447) & 0.0698 \\
& $Q=3$ & 0.97 & 0.0608 (0.0362) & 0.0764 && 0.96 & 0.0788 (0.0371) & 0.0723 \\
& $Q=5$ & 1.00 & 0.0488 (0.0364) & 0.0786 && 0.99 & 0.0575 (0.0370) & 0.0743 \\ 
& $Q=7$ & 1.00 & 0.0400 (0.0293) & 0.0772 && 1.00 & 0.0289 (0.0226) & 0.0751 \\
& \underline{Matching} & 0.96 & 0.0499 (0.0392) & 0.0715 && 0.97 & 0.0454 (0.0348) & 0.0649\\
& \underline{IPW} & 0.99 & 0.0403 (0.0288) & 0.0600 && 0.96 & 0.0362 (0.0294) & 0.0531\\
\midrule
0.50 & \underline{ABB} & & & & & & & \\
& $Q=1$ & 0.64 & 0.1013 (0.0426) & 0.0596 && 0.02 & 0.1818 (0.0391) & 0.0536 \\
& $Q=3$ & 0.85 & 0.0731 (0.0435) & 0.0680 && 0.76 & 0.0929 (0.0399) & 0.0607 \\
& $Q=5$ & 1.00 & 0.0403 (0.0317) & 0.0757 && 0.98 & 0.0455 (0.0331) & 0.0670 \\
& $Q=7$ & 1.00 & 0.0447 (0.0352) & 0.0789 && 1.00 & 0.0375 (0.0283) & 0.0725 \\
& \underline{Matching} & 0.94 & 0.0784 (0.0612) & 0.0839 && 0.91 & 0.0596 (0.0493) & 0.0689\\
& \underline{IPW} & 0.94 & 0.2321 (0.3459) & 0.1266 && 0.92 & 0.2223 (0.3213) & 0.1508\\
\midrule
0.75 & \underline{ABB} & & & & & & & \\
& $Q=1$ & 0.48 & 0.0955 (0.0350) & 0.0464 && 0.03 & 0.1538 (0.0316) & 0.0410 \\
& $Q=3$ & 0.97 & 0.0441 (0.0330) & 0.0636 && 0.86 & 0.0600 (0.0374) & 0.0545 \\
& $Q=5$ & 0.98 & 0.0661 (0.0474) & 0.0896 && 0.99 & 0.0464 (0.0328) & 0.0696 \\
& $Q=7$ & 0.94 & 0.0808 (0.0598) & 0.0932 && 0.95 & 0.0601 (0.0462) & 0.0808 \\
& \underline{Matching} & 0.88 & 0.1355 (0.1099) & 0.1153 && 0.82 & 0.1055 (0.0851) & 0.0904\\
& \underline{IPW} & 0.60 & 0.3117 (0.4725) & 0.1092 && 0.65 & 0.7027 (0.8125) & 0.2890\\
\midrule
1.00 & \underline{ABB} & & & & & & & \\
& $Q=1$ & 0.42 & 0.0749 (0.0257) & 0.0376 && 0.01 & 0.1169 (0.0251) & 0.0310 \\
& $Q=3$ & 0.97 & 0.0487 (0.0444) & 0.0696 && 0.95 & 0.0414 (0.0326) & 0.0543 \\
& $Q=5$ & 0.98 & 0.0901 (0.0667) & 0.1026 && 0.99 & 0.0448 (0.0334) & 0.0760 \\
& $Q=7$ & 0.87 & 0.1428 (0.0981) & 0.1208 && 0.94 & 0.1095 (0.0854) & 0.1086 \\
& \underline{Matching} & 0.94 & 0.1546 (0.1349) & 0.1648 && 0.83 & 0.1547 (0.1437) & 0.1297\\
& \underline{IPW} & 0.61 & 0.4113 (0.5380) & 0.1311 && 0.65 & 0.3264 (0.4125) & 0.1414\\
\bottomrule
\end{tabular}
\label{sim.ordinal}
\end{sidewaystable}

\newpage

\begin{table}[!ht]
\centering
\caption{Estimated 3-year rates (expressed as \%) for MACE and ACM (with 95\% confidence interval) among gliclazide second-line users}
\begin{tabular}{l l l l l l}
\toprule
Outcome & Method && Gliclazide & Sitagliptin & Pioglitazone \\
\midrule
\underline{MACE} & \underline{ABB} && & &\\
& $Q=1$ && 12.61 (11.97, 13.24) & 11.04 (9.58, 12.49) & 7.96 (6.61, 9.32)\\
& $Q=3$ && 12.61 (11.97, 13.24) & 12.54 (11.03, 14.05) & 8.23 (6.61, 9.86)\\
& $Q=5$ && 12.61 (11.97, 13.24) & 12.46 (10.80, 14.12) & 8.92 (7.30, 10.55)\\
& $Q=7$ && 12.61 (11.97, 13.24) & 12.55 (10.83, 14.27) & 9.11 (7.52, 10.70)\\
\midrule\midrule
\underline{ACM} & \underline{ABB} && & &\\
& $Q=1$ && 3.87 (3.50, 4.24) & 2.26 (1.57, 2.94) & 3.19 (2.33, 4.04)\\
& $Q=3$ && 3.87 (3.50, 4.24) & 2.65 (1.99, 3.32) & 3.43 (2.28, 4.58)\\
& $Q=5$ && 3.87 (3.50, 4.24) & 2.72 (1.84, 3.60) & 3.54 (2.61, 4.48)\\
& $Q=7$ && 3.87 (3.50, 4.24) & 2.78 (2.00, 3.56) & 3.39 (2.43, 4.35)\\
\bottomrule
\end{tabular}
\label{baseline.table}
\end{table}

\begin{table}[!ht]
\centering
\caption{Estimated 3-year risk ratio for MACE and ACM (95\% confidence interval) among gliclazide second-line users}
\begin{tabular}{l l l l l }
\toprule
Outcome & Method && Sitagliptin vs. Gliclazide & Pioglitazone vs. Gliclazide \\
\midrule
\underline{MACE} & \underline{ABB} && & \\
& $Q=1$ && 0.88 (0.77, 1.00)$^{*}$ & 0.63 (0.56, 0.72)$^{*}$ \\
& $Q=3$ && 0.97 (0.85, 1.11) &  0.70 (0.60, 0.82)$^{*}$\\
& $Q=5$ && 0.96 (0.85, 1.08) & 0.69 (0.59, 0.80)$^{*}$ \\
& $Q=7$ && 0.99 (0.86, 1.15) & 0.72 (0.63, 0.83)$^{*}$ \\
\midrule\midrule
\underline{ACM} & \underline{ABB} && & \\
& $Q=1$ && 0.58 (0.47, 0.73)$^{*}$ & 0.82 (0.64, 1.06)\\
& $Q=3$ &&0.70 (0.55, 0.87)$^{*}$ & 0.89 (0.67, 1.18) \\
& $Q=5$ && 0.71 (0.54, 0.93)$^{*}$ & 0.89 (0.67, 1.17) \\
& $Q=7$ && 0.72 (0.57, 0.91)$^{*}$ & 0.87 (0.67, 1.15) \\
\bottomrule
\end{tabular}
\label{relative.risk.table}
\end{table}

\end{document}